\documentclass[%
 reprint,
superscriptaddress,
showpacs,
 amsmath,amssymb,
pra,
]{revtex4-1}

\usepackage{graphicx}
\usepackage{dcolumn}
\usepackage{bm}
\usepackage{color}
\usepackage{hyperref}
\hypersetup{backref,pdfpagemode=FullScreen,colorlinks=true,breaklinks,urlcolor=blue,linkcolor=blue,citecolor=blue}
\usepackage{siunitx}

\begin{document}

\preprint{APS/123-QED}

\title{Observation of broad $p$-wave Feshbach resonances in ultracold $^{85}$Rb-$^{87}$Rb mixtures}
\author{Shen Dong}
\author{Yue Cui}
\author{Chuyang Shen}
\author{Yewei Wu}
\thanks{Present address: JILA, University of Colorado at Boulder, Boulder, CO 80309.}
\affiliation{State Key Laboratory of Low Dimensional Quantum Physics, Department of Physics, Tsinghua University, Beijing 100084, China}
\author{Meng Khoon Tey}
\email{mengkhoon\_tey@mail.tsinghua.edu.cn}
\affiliation{State Key Laboratory of Low Dimensional Quantum Physics, Department of Physics, Tsinghua University, Beijing 100084, China}
\affiliation{Collaborative Innovation Center of Quantum Matter, Beijing, China}
\author{Li You}
\email{lyou@mail.tsinghua.edu.cn}
\affiliation{State Key Laboratory of Low Dimensional Quantum Physics, Department of Physics, Tsinghua University, Beijing 100084, China}
\affiliation{Collaborative Innovation Center of Quantum Matter, Beijing, China}
\author{Bo Gao}
\email{bo.gao@utoledo.edu}
\affiliation{Department of Physics and Astronomy, University of Toledo, Mailstop 111, Toledo, Ohio 43606, USA}


\date{\today}

\begin{abstract}

We observe new Feshbach resonances in ultracold mixtures of $^{85}$Rb and $^{87}$Rb atoms in the $^{85}$Rb$|2, +2\rangle$+$^{87}$Rb$|1, +1\rangle$ and $^{85}$Rb$|2, -2\rangle$+$^{87}$Rb$|1, -1\rangle$ scattering channels. The positions and properties of the resonances are predicted and characterized using the semi-analytic multichannel quantum-defect theory by Gao. Of particular interest, a number of broad entrance-channel dominated $p$-wave resonances are identified, implicating exciting opportunities for studying a variety of $p$-wave interaction dominated physics.

\end{abstract}

\maketitle

\section{Introduction}

Feshbach resonance (FR) is a powerful toolbox in the field of ultracold atoms~\cite{chin2010feshbach}. It occurs when the energy of two scattering atoms becomes quasi-degenerate with their molecular state in a different spin configuration. Experimentally, one can tune two atoms into a FR by changing magnetic field taking advantage of the differential magnetic dipole moment between the atomic state and the molecular state. Magnetically-induced FRs have found tremendous applications. They have been used, for instance, to control scattering and loss properties of cold atomic gases~\cite{chin2010feshbach,Cornish2000TuningInteractions,Strecker2002Li7,Weber2003CsBEC}, to study universal few-body physics such as Efimov states~\cite{chin2010feshbach,Kraemer2006,vonStecher2009,Jochim2013FewAtoms}, to create Feshbach or ro-vibrational ground state  molecules~\cite{regal2003creation,ni2008high,lang2008ultracold,danzl2010ultracold}, and to study BEC-BCS crossover and Fermi gases in the unitarity limit~\cite{Kinast1296,Horikoshi442,Nascimbene2010,Ku2012EOS,Sidorenkov2013,Tey2013CollectiveModes,Edmundo2013HigherOrder}.

Many of the aforementioned applications make use of ``broad" $s$-wave FRs whose properties are dominated by the entrance (open) channel. These resonances acquire the universal behavior demonstrated by a single-channel resonance and therefore are useful for studying universal properties in few-body and many-body atomic systems. There exist a number of ``broad" $s$-wave FRs in ultracold alkali-metal systems, for both intraspecies~\cite{chin2000high,PhysRevLett.89.203201,PhysRevLett.88.173201,knoop2011feshbach,blackley2013feshbach} and interspecies~\cite{papp2006observation,ferlaino2006feshbach,wille2008exploring,deh2008feshbach, repp2013observation,cho2013feshbach, wang2013observation} mixtures. However, it is not clear if ``broad" FRs also exist for higher partial waves. This is because the criterion for distinguishing ``broad" and ``narrow" non-$s$-wave FRs was not established firmly until the multichannel quantum-defect theory (MQDT) is extended to neutral atom with van der Waals interaction~\cite{Gao2005MQDT, gao2011analytic}. This theory enables a uniform approach to treat FRs in higher partial waves as in the $s$-wave, it can therefore provide a rigorous definition of ``broad" FRs for all partial waves~\cite{gao2011analytic}. The first application of this model to the mixture of $^6$Li and $^{40}$K~\cite{makrides2014multichannel} found all resonances, both $s$-wave and higher partial waves, to be closed-channel dominated and therefore are all ``narrow" resonances.

In this paper, we apply the semi-analytic MQDT~\cite{gao2011analytic,makrides2014multichannel} to predict and characterize FRs in different spin and partial-wave channels of the $^{85}$Rb-$^{87}$Rb mixtures. We find that this mixture exhibits a rich spectrum of ``broad" $s$-wave and $p$-wave FRs. The predicted FRs are observed and verified experimentally. In particular, a very broad entrance-channel dominated $p$-wave resonance we discovered in the lowest energy channel of the system suggests exciting opportunities for investigating universal few-body and many-body behaviors with strong coupling in nonzero partial waves~\cite{PhysRevLett.90.053202,PhysRevLett.95.230401,PhysRevLett.98.200403,Ohashi2005BECBCS,Jiang2014pwave,Thywissen2016pwaveContact}, including $p$-wave bosonic superfluid mixture, three-body recombination decay, and formation of $p$-wave heteronuclear molecules.

\section{``Broad" or ``Narrow"}

Besides theoretical curiosity, a main motivation for distinguishing whether a FR is intrinsically ``broad" or ``narrow" is to determine its usefulness for studying universal properties of strongly interacting few-body and many-body atomic systems at low temperatures. For cold neutral atoms in the ground states, such universal properties arise from the common long-range $-1/R^6$ type ($R$ being the interatomic separation) attractive van der Waals potential between the atoms, and the fact that the low-temperature near-threshold scattering wave functions of these systems are predominantly shaped by the long-range potentials~\cite{chin2010feshbach,Gao2000universal}. Since collision of real atoms is intrinsically a multichannel process, systems near a FR (which provides the required strong interactions) might or might not exhibit universal behaviors expected for single-channel collisions. This explains the interest in identifying ``broad" FRs.

A ``broad" FR is technically defined as one that shows universal characteristics of a single-channel collision over a large fraction of its resonance width, where the overall wave function of the colliding atoms is dominated by that from the entrance channel over that from the quasi-degenerate closed channel. A nice and detailed discussion about the characteristics of ``broad" $s$-wave FRs can be found in Ref.~\cite{chin2010feshbach}. Here, we extend the discussion to include higher partial waves from the perspective of the semi-analytic MQDT.

\begin{figure}
    \includegraphics[width=0.9\linewidth]{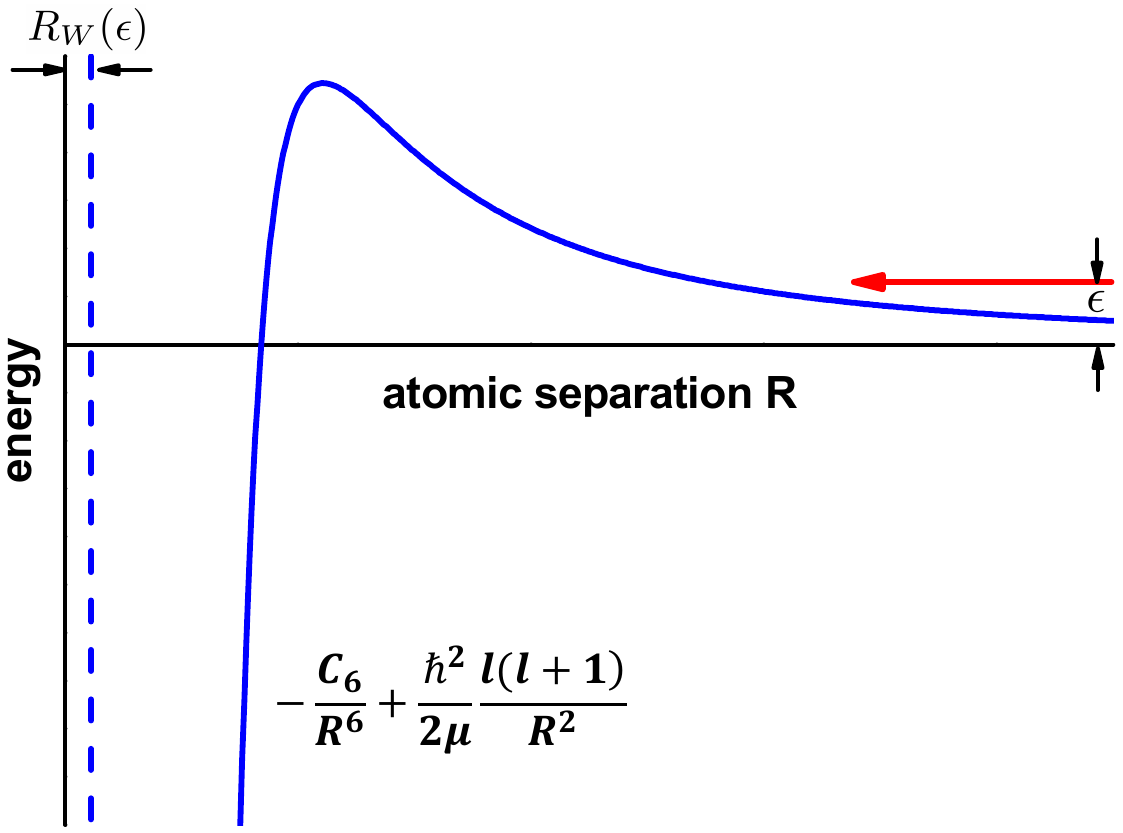}
    \caption{(Color online) The essence of MQDT based on the analytic solutions to the Schr\"odinger equation with a $-1/R^6$ potential~\cite{gao2011analytic}. The real molecular potentials are replaced by a single pure $-\frac{C_6}{R^6}+\frac{\hbar^2}{2\mu}\frac{l(l+1)}{R^2}$ potential (extended from the real long-range potential) cut off by a sharp infinite repulsive wall (dashed line) at $R_W$ which accounts for the multichannel effect from short ranges. The effective ``quantum defect" $R_W(\epsilon)$ depends sensitively on the scattering energy $\epsilon$ for a ``narrow" FR.}
    \label{broadOrNarrow}
\end{figure}

Considering two-atom scattering with only one channel, the physics of the system near the threshold energy is almost independent of the details of the short-range interactions. It can be fully determined by the long-range potential and a quantum-defect parameter which takes in the effect of  the short-range interactions. This quantum defect depends weakly on the energy and the partial wave quantum number $l$ of the system under study. For the quantum-defect theory (QDT) based on the analytic solutions to the Schr\"odinger equation with a $-1/R^6$ potential~\cite{gao1998r6solutions,Gao2005MQDT,gao2011analytic}, the near-threshold physics of the system can be studied most efficiently by replacing the real molecular potential by a pure $-\frac{C_6}{R^6}+\frac{\hbar^2}{2\mu}\frac{l(l+1)}{R^2}$ potential which is cut off by a sharp infinite repulsive wall at $R_W$ at a sufficiently small $R$ (see Fig.~\ref{broadOrNarrow} for illustration). Here, the $\frac{\hbar^2}{2\mu}\frac{l(l+1)}{R^2}$ term represents the centrifugal barrier of the $l$-th partial wave and $C_6$ is the actual van der Waals interaction coefficient for the system. The infinite wall sets the boundary conditions and defines the relative phase between the incoming and outgoing wave functions at small $R$, effectively replacing the short-range potential. In QDT, its effect is represented by the short-range $K$ matrix ($K^c$ ) or its equivalent - the quantum defect $\mu^c$~\cite{Fano1970,Starace1976,Seaton1983,Gao2008General1overRn}.

For a real atomic system with many scattering channels, if the collision energy lies between the ground channel and the threshold for the first higher energy channel, the scattering problem can be reduced to a single-channel using an effective short-range $K^c_\mathrm{eff}$ matrix which takes into account the influence of all channels~\cite{Mies1984MQDT,Gao2005MQDT,gao2011analytic,Idzaszek2011MQDT}. In this scenario (MQDT), the equivalent parameter $R_W(\epsilon)$ becomes more dependent on the energy $\epsilon$ (Fig.~\ref{broadOrNarrow}) than the single-channel case. For a ``broad" FR, the energy dependence of $K^c_\mathrm{eff}$ or $R_W(\epsilon)$ should exert insignificant influence on the properties of the system (as a function of $\epsilon$) that are determined by the long-range van der Waals interaction only.

\section{Predicting and characterizing Feshbach resonances of all partial waves}
We use the semi-analytic MQDT~\cite{gao2011analytic,makrides2014multichannel} to predict and characterize the FRs for various partial waves in the $^{85}$Rb-$^{87}$Rb mixture. The model we adopt for the current study leaves out the weak magnetic dipole-dipole~\cite{Moerdijk1995SpinExchange,Stoof1988SpinExchange} as well as the second-order electronic spin-orbit interactions~\cite{Mies1996Collisions,Kotochigova2000SecondOrderSpinOrbit,Julienne2000FBCs}, therefore it only predicts FRs due to couplings between atomic and molecular states of the same partial wave. These FRs are typically broader and thus easier to observe. To set the benchmark for future investigations, we limit ourselves to the baseline MQDT results that ignore the weak energy and partial-wave dependence of the quantum defects, i.e., we adopt the approximations of $K_{S}^{c}(\epsilon,l)\approx K_{S}^{c}(\epsilon = 0,l = 0)$ and $K_{T}^{c}(\epsilon,l)\approx K_{T}^{c}(\epsilon = 0,l = 0)$~\cite{gao2011analytic} for the singlet- and triplet-state short-range $K$ matrices, respectively. In this baseline description, all aspects of cold atomic interaction, including parameters for all magnetic FRs in all partial waves, are determined from three parameters~\cite{Gao2005MQDT, PhysRevA.79.040701}: the $C_{6}$ coefficient, the singlet $s$-wave scattering length $a_{l = 0}^{S}$, and the triplet $s$-wave scattering length $a_{l = 0}^{T}$, apart from the well-known atomic parameters such as the atomic mass and hyperfine splitting.

The three parameters we adopt for this study are $C_{6}$ = 4710 a.u., $a_{l = 0}^{S}$ = 11.37 a.u., and $a_{l = 0}^{T}$ = 184.0 a.u. The $C_{6}$ coefficient and the singlet $s$-wave scattering length are taken unaltered from Ref.~\cite{PhysRevA.82.052514}, while the triplet $s$-wave scattering length is adjusted from the value of 201.0 in Ref.~\cite{PhysRevA.82.052514} to 184.0 a.u. such that it agrees better with experimental $s$-wave FR positions reported in Ref.~\cite{papp2006observation}. The above scattering lengths we adopt correspond to quantum defects of $\mu_{S}^{c}(\epsilon = 0,l = 0)$ = 0.7253 for the singlet state and $\mu_{T}^{c}(\epsilon = 0,l = 0)$ = 0.2045 for the triplet state~\cite{Gao2008General1overRn}. Their respective short-range $K$ matrices are $K_{S}^{c}(\epsilon,l)\approx K_{S}^{c}(\epsilon = 0,l = 0) = -0.5084$ and $K_{T}^{c}(\epsilon,l)\approx K_{T}^{c}(\epsilon = 0,l = 0) = 1.685$~\cite{Gao2005MQDT, Gao2008General1overRn}, from which the $K^c$ matrix in a magnetic field is constructed, and predictions and characterizations of FRs are carried out as discussed in Ref.~\cite{makrides2014multichannel}.

\begin{figure*}
    \includegraphics[width=0.7\linewidth]{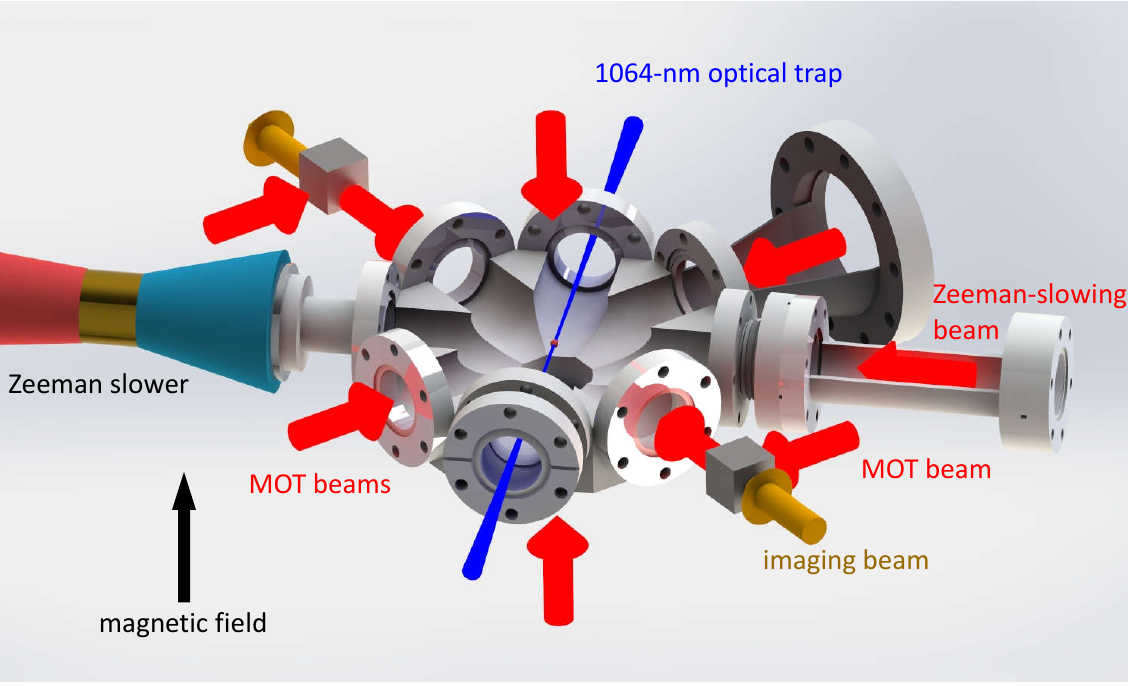}
    \caption{(Color online) The essential elements of our experimental setup. $^{87}$Rb and $^{85}$Rb atoms are loaded into a magneto-optical trap from a common Zeeman slower. A 1064-nm optical dipole trap is used to confine the atoms for FR loss spectroscopy. A single bichromatic imaging beam overlapping with one of the MOT beams is used to probe the atom numbers of both species in sequence.}
    \label{exptSetup}
\end{figure*}

According to the formulation and  discussions in Ref.~\cite{gao2011analytic}, around each resonance located at $B_{0l}$, especially an isolated resonance, the dependence of the generalized scattering length on the magnetic field can be parameterized for all partial waves by
\begin{equation}\label{scattEqn}
\tilde{a}_{l}(B) = \tilde{a}_{\textmd{bgl}}\left(1 - \frac{\Delta_{Bl}}{B-B_{0l}}\right),
\end{equation}
similar to the usual form for the $s$-wave FRs~\cite{chin2010feshbach}. Here, $\tilde{a}_{\textmd{bgl}}$ (of dimension L$^{2l+1}$) is the generalized background scattering length for the $l$-th partial wave~\cite{gao2009analytic} and $|\Delta_{Bl}|$ denotes the distance from the FR position to the nearest zero-$\tilde{a}_{l}$ position. These parameters, $B_{0l}$, $\tilde{a}_{\textmd{bgl}}$ and $\Delta_{Bl}$, together with $\delta\mu_{l}$ (the differential magnetic moment between the molecular state and the atomic state)~\cite{chin2010feshbach, gao2011analytic} and the $C_{6}$ coefficient, form a set of parameters that provides a complete description of atomic interaction around a magnetic FR. They constitute the most direct generalization of the $s$-wave description \cite{chin2010feshbach,Moerdijk1995SpinExchange,Julienne2006RMP} to higher partial waves. They are easily adapted and highly efficient for treating FR at zero energy (the threshold).

A derived parameter which distinguishes a ``broad" resonance ($|\zeta_\textrm{res}|\gg1$) that follows single-channel universal behaviors from a ``narrow" resonance ($|\zeta_\textrm{res}|\ll1$) that violates such a behavior is given by~\cite{gao2011analytic}
\begin{equation}\label{zeta}
\zeta_{\textmd{res}} = -\frac{1}{(2l+3)(2l-1)} \frac{\tilde{a}_{\textmd{bgl}}}{\bar{a}_{l}} \left( \frac{\delta\mu_{l} \Delta_{Bl}}{s_{E}} \right).
\end{equation}
When $l=0$, the parameter $\zeta_{\textrm{res}}$ reduces to the $s$-wave dimensionless resonance strength parameter $s_{\textrm{res}}$ previously introduced \cite{chin2010feshbach} except for a factor of 3. In Eq.~(\ref{zeta}), $s_E=\hbar^2/(2\mu\beta^2_6)$ is the characteristic energy scale of the van der Waals potential, with $\beta_6=\left(2\mu C_6/\hbar^2\right)^{1/4}$ being the characteristic length scale and $\mu$ the reduced mass. The parameter $\bar{a}_{l}=\bar{a}_{sl}\beta_6^{2l+1}$ is the mean scattering length for the $l$-th partial wave, with~\cite{gao2011analytic}
\begin{equation}
\bar{a}_{sl}=\frac{\pi^2}{2^{4l+1}\left[\Gamma(l/2+1/4)\Gamma(l+3/2)\right]^2}.
\end{equation}

\begin{figure*}
    \includegraphics[width=1\linewidth]{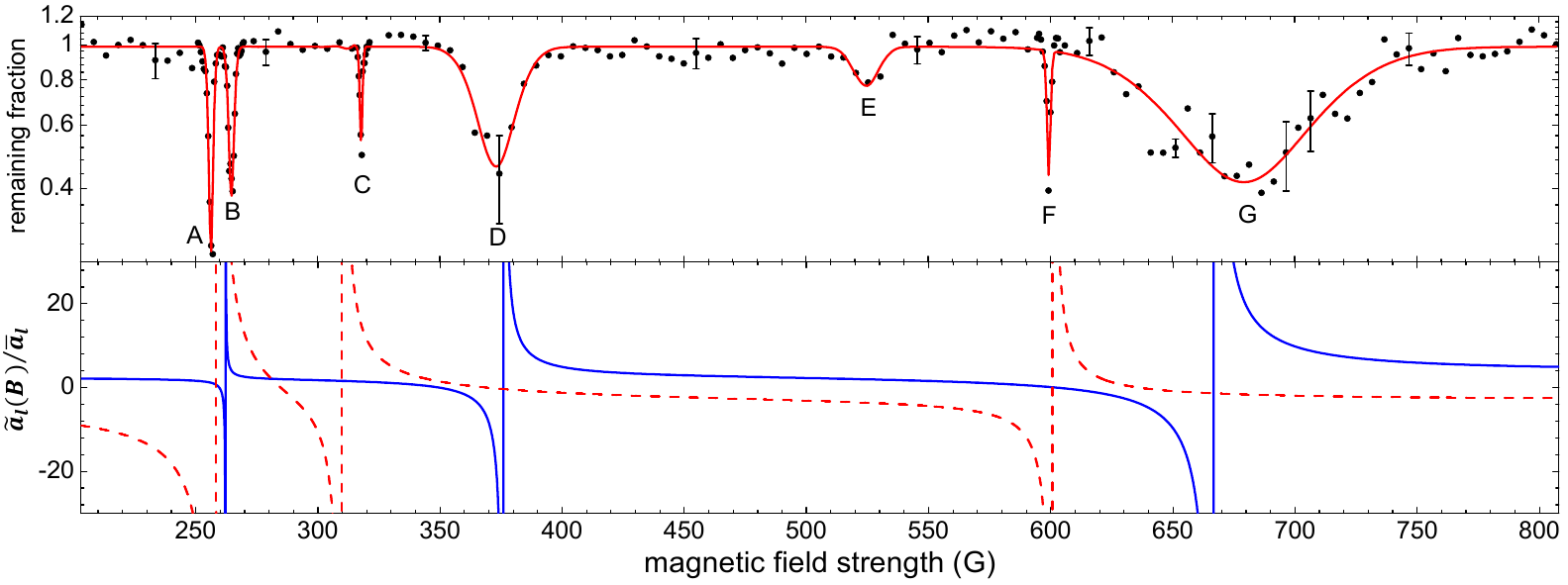}
    \caption{(Color online) Feshbach spectrum for the $^{85}$Rb$|2,-2\rangle + ^{87}$Rb$|1,-1\rangle$ channel. Top panel: The remaining fraction of $^{85}$Rb atoms normalized to the off-resonant values after 400-1000\,ms interaction with $^{87}$Rb at $\sim$\SI{2}{\micro K} is shown as a function of magnetic field. Every data point is averaged over 5 runs. Error bars denoting typical standard deviations are shown at specific magnetic fields for illustrative purposes. We fit the data using a multi-peak Gaussian profile to get the eye-guiding line (red solid line). Bottom panel: The $s$-wave (blue solid line) and $p$-wave (red dashed line) RGSL, $\tilde{a}_l(B)/\bar{a}_l$, computed using the semi-analytic MQDT.
    }
    \label{MFm3GrossScan}
\end{figure*}

\section{Experiment}
Figure~\ref{exptSetup} illustrates the essential elements of our experimental setup. Both $^{87}$Rb and $^{85}$Rb atoms are first loaded into a magneto-optical trap (MOT) from a common Zeeman slower. The $^{87}$Rb atoms are loaded for 8\,s while the $^{85}$Rb atoms are only loaded simultaneously during the last 1\,s of the $^{87}$Rb loading stage. This procedure gives about $2.5\times10^{9}$ $^{87}$Rb atoms and $5\times10^{7}$ $^{85}$Rb atoms following the molasses cooling stage. After optical pumping both isotopes to their low-field seeking hyperfine states, $^{85}$Rb$|f=2,m_f=-2\rangle$ and $^{87}$Rb$|1, -1\rangle$, the atoms are loaded into a magnetic quadrupole trap. Forced microwave evaporation is then performed on $^{87}$Rb atoms, and $^{85}$Rb atoms are sympathetically cooled by the $^{87}$Rb atoms to reach a common temperature of $\sim$\SI{12}{\micro\K}. The mixture is then loaded into an optical dipole trap formed by a single focused  horizontal 1064-nm light beam, which has a $1/e^{2}$ waist of $\sim$\SI{35}{\micro\meter} and an initial power of 2.7\,W. At this stage, we typically have $3.5\times10^{6}$ $^{87}$Rb atoms and $4\times10^{6}$ $^{85}$Rb atoms. By reducing the light power from 2.7\,W to 0.25\,W in 3.2\,s in two linear ramps, the mixture is further cooled down to a temperature of $\sim$\SI{2}{\micro\K} and ends up with approximately $1.5\times10^{6}$ $^{87}$Rb atoms and $5\times10^{5}$ $^{85}$Rb atoms. At this stage, the trapping frequencies are $\approx 2\pi\times$ (2.5, 376, 374)\,Hz for $^{87}$Rb atoms.

The remaining $^{85}$Rb and $^{87}$Rb atoms are then prepared into the desired spin states (scattering channels). Afterwards, we switch the magnetic field to a certain value in the range of 0-900\,G and then hold the mixture for some amount of time to search for the FRs predicted by theory. The presence of FRs results in enhanced atomic scattering and three-body collision loss, and thereby shows up as loss features in the loss spectrum as a function of the magnetic field. We determine the atom numbers of the two isotopes by absorption imaging using a CCD camera working in the fast-kinetic mode. The magnetic field strength is calibrated by measuring the Zeeman splittings of $^{87}$Rb magnetic sublevels which are then compared to the Breit-Rabi formula. The magnetic-field fluctuation in our system is typically about 150\,mG, mainly limited by variation in the magnetization of the stainless-steel vacuum chamber in different experimental cycles.

\section{Results}
Figure~\ref{MFm3GrossScan} shows the $s$-wave and $p$-wave FRs for the $^{85}$Rb$|2,-2\rangle+^{87}$Rb$|1,-1\rangle$ scattering channel from 200\,G to 800\,G. The top panel displays the remaining $^{85}$Rb fraction relative to the off-resonant values after co-existing with $^{87}$Rb for typical durations from 400\,ms to 1000\,ms as a function of magnetic field. The bottom panel shows the predictions of the reduced generalized scattering length (RGSL), $\tilde{a}_l(B)/\bar{a}_l$, for both the $s$-wave (blue solid line) and $p$-wave (red dashed line) channels from the semi-analytic MQDT~\cite{gao2009analytic,gao2011analytic,makrides2014multichannel}. It is clear from the theoretical results that the features at B, D, and G correspond to the $s$-wave FRs, while those at A, C, and F are $p$-wave FRs. The loss at E, which is present even without $^{87}$Rb atoms, is due to an intraspecies FR of $^{85}$Rb itself~\cite{blackley2013feshbach}. Among the resonances observed, A, B, D were reported previously in Ref.~\cite{papp2006observation}, and C was also predicted in Ref.~\cite{ticknor2004multiplet}.

\begin{figure*}
    \includegraphics[width=1\linewidth]{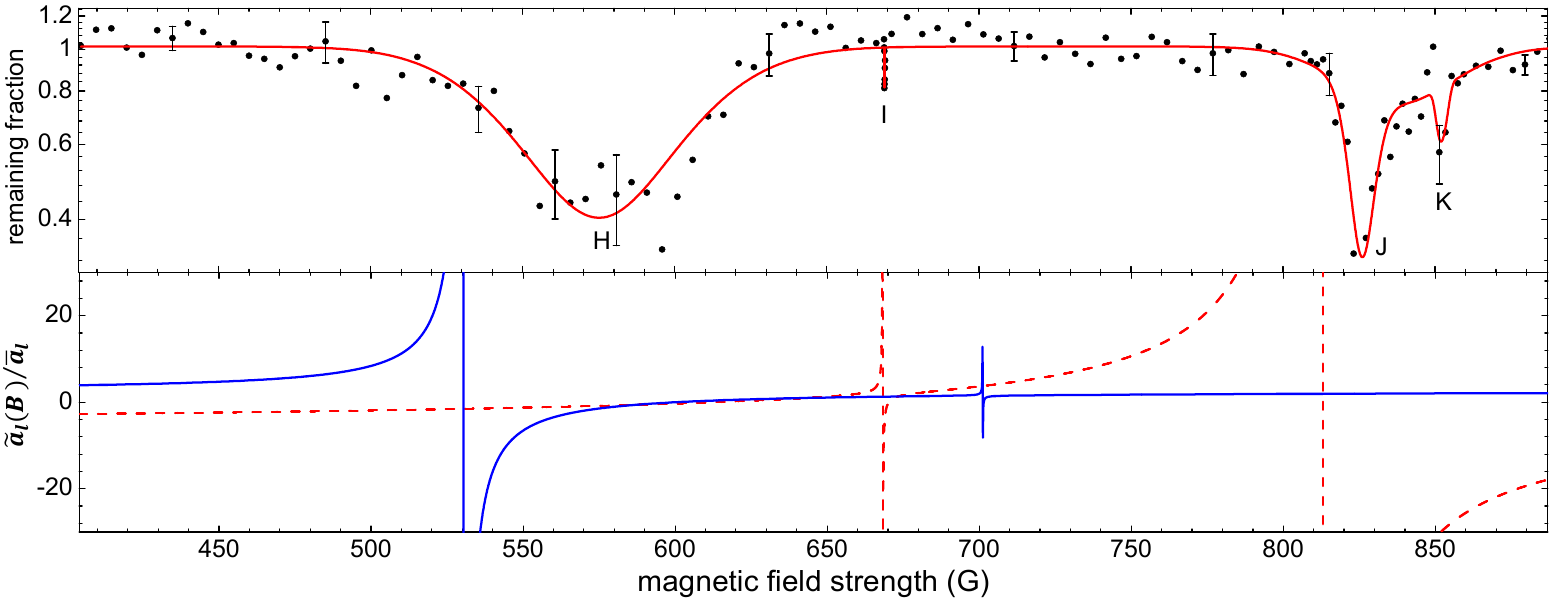}
    \caption{(Color online) The same as in Fig.~\ref{MFm3GrossScan} but for the $^{85}$Rb$|2,+2\rangle + ^{87}$Rb$|1,+1\rangle$ channel after 600-ms interaction time. }
    \label{MF3GrossScan}
\end{figure*}

Figure~\ref{MF3GrossScan} displays the FRs for the $^{85}$Rb$|2,+2\rangle$+ $^{87}$Rb$|1,+1\rangle$ scattering channel which has not been reported before. After preparing the mixture at $\sim$\SI{2}{\micro\K}, the $^{85}$Rb and $^{87}$Rb atoms are transferred to the $|2, +2\rangle$ and $|1, +1\rangle$ states, respectively, by rf adiabatic passages~\cite{peik1997bloch}. Our model predicts two $s$-wave and two $p$-wave FRs for this channel from 0 to 900\,G. We find three broader FRs but not the narrowest one predicted at around 701\,G. These FRs are labeled by H, I, and J in Fig.~\ref{MF3GrossScan}. The resonance at K is again due to an intraspecies FR of the $^{85}$Rb atoms~\cite{blackley2013feshbach}. In these measurements, the scanning step size was set at 5\,G for most magnetic field range as most of the FRs we are interested in are quite broad. As a result, the measurements could miss narrower FRs such as those theoretically predicted for the $d$ or even higher partial waves, those due to coupling between atomic and molecular channels of different partial waves, and those caused by $^{87}$Rb or $^{85}$Rb intraspecies scattering~\cite{PhysRevLett.89.283202,blackley2013feshbach}. These not observed narrow FRs are not the intended targets for the current study.

\begin{figure}
    \includegraphics[width=0.9\linewidth]{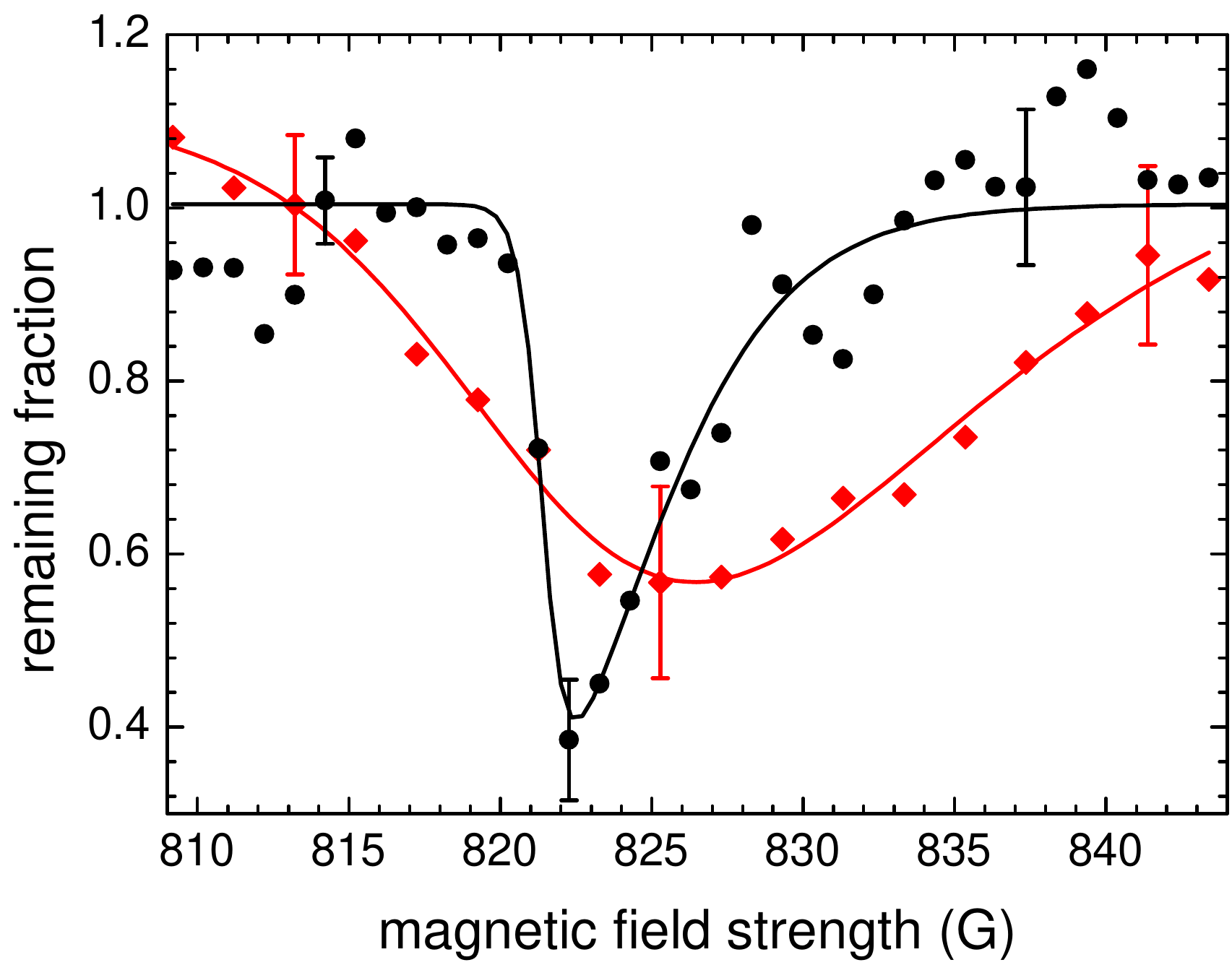}
    \caption{(Color online) Temperature dependence of the broadest $p$-wave resonance observed in the $^{85}$Rb$|2,+2\rangle+^{87}$Rb$|1,+1\rangle$ channel (J in Fig.~\ref{MF3GrossScan}). The loss feature narrows, and its left-edge becomes steeper as the temperature of the mixture is lowered from $\sim$\SI{2}{\micro K} (red diamonds) to $\sim$400\,nK (black circles). The solid lines represent the fits to the data using asymmetric double sigmoid function to account for the asymmetric line shapes. }
    \label{broadestP}
\end{figure}

\begin{figure}
    \includegraphics[width=0.9\linewidth]{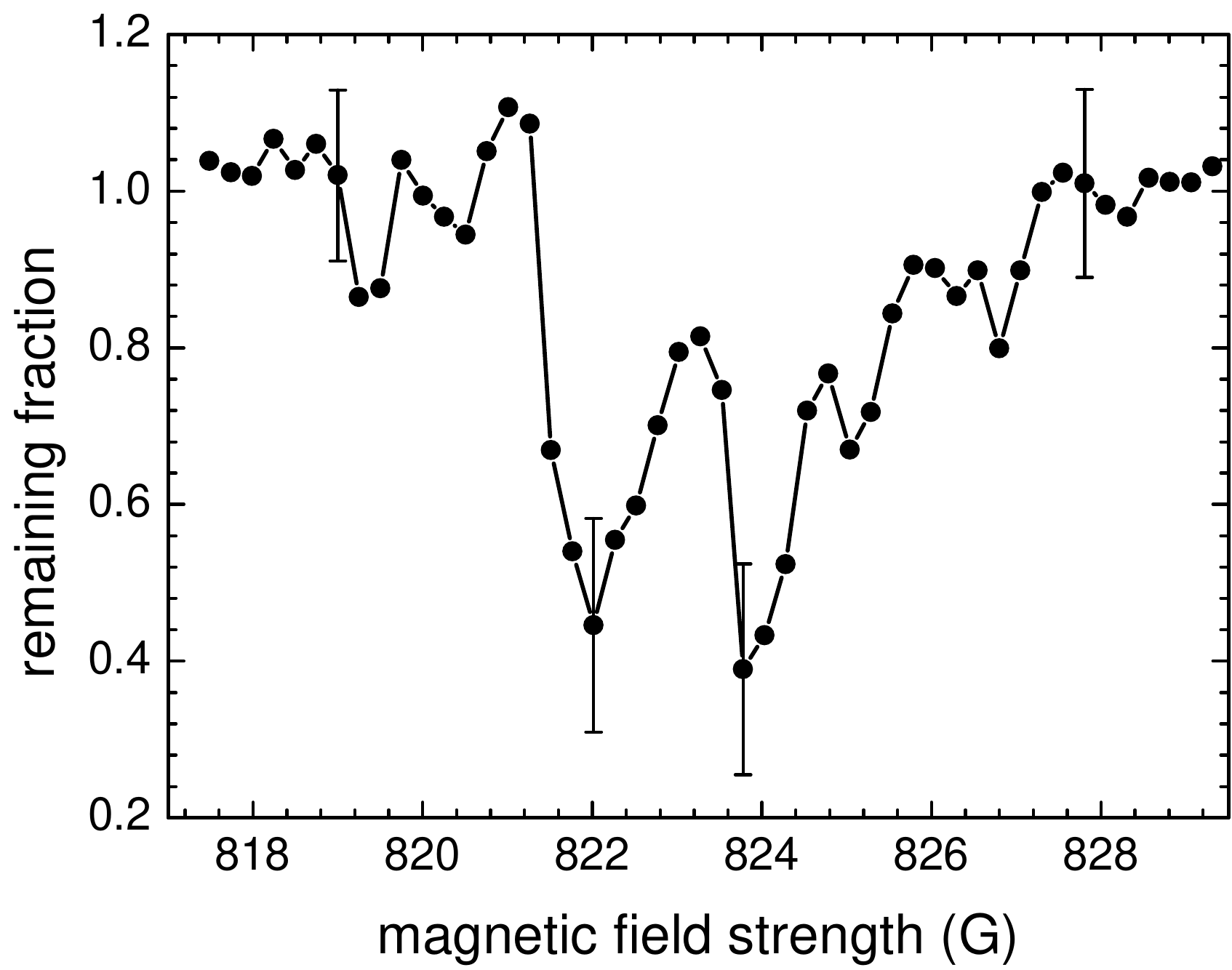}
    \caption{A clear doublet splitting emerges for the 823-G $p$-wave FR in Fig.~\ref{MF3GrossScan} at a temperature of  $\sim$260\,nK.}
    \label{Pdoublet}
\end{figure}

We identify the $p$-wave resonances not only by their proximity to our theoretical predictions, but also by their asymmetric line shapes and more definitely by their distinctive doublet structures~\cite{ticknor2004multiplet} which can be observed when their loss-feature widths become smaller than their doublet splitting at sufficiently low temperatures. Fig.~\ref{broadestP} shows a distinctive change in the line shape of the broadest $p$-wave FR (J in Fig.~\ref{MF3GrossScan}) when the temperature of the mixture is lowered from $\sim$\SI{2}{\micro K} (red diamonds) to $\sim$400\,nK (black circles). As the collision energy gets nearer to the channel's threshold, the loss feature becomes narrower and its `threshold' edge becomes steeper. These observed features are in agreement with the expectations for a high partial wave resonance~\cite{ticknor2004multiplet}. Fig.~\ref{Pdoublet} shows the doublet structure of the same $p$-wave resonance observed at a temperature of $\sim$260\,nK. Here, an extra 1064-nm light beam with a $1/e^{2}$ waist of $\sim$\SI{120}{\micro\meter} was added to form a crossed optical dipole trap in order to enhance the collision rate. We have observed the doublet structures of all p-wave FRs reported here except for the narrowest one at 669.0(2)\,G. Nevertheless, we can confirm that this FR is non-$s$-wave since its loss feature disappears at sufficiently low temperatures.

The experimentally measured positions and widths of the FR loss features are tabulated in Table~\ref{table}, together with the theoretical predictions for the relevant resonance parameters, namely $B_{0l}$, $\tilde{a}_{\textmd{bgl}}$, $\Delta_{Bl}$, $\delta\mu_{l}$ and $\zeta_{\textmd{res}}$. The resonance centers and full-width-at-half-maximum (FWHM) of all $s$-wave loss features are obtained using Gaussian fit. The positions of the $p$-wave resonances are determined by the intercept of the lossless baseline with the steeper edge of the $p$-wave loss features, while the reported widths are obtained from Gaussian fit. The reported uncertainties in the measured FR positions $B_{0l}^{\textmd{expt}}$ represent the loss-feature fitting uncertainties. For fitting uncertainties smaller than 0.2\,G, they are set to 0.2\,G to reflect the amplitude of magnetic field fluctuation in our experimental setup. The only quantity in Table~\ref{table} that is directly comparable between theory and experiment is the resonance position, for which the agreements are within a few percent. They are better for the $^{85}$Rb$|2,-2\rangle+^{87}$Rb$|1,-1\rangle$ channel because we had adopted the best fitted $a^T_{l=0}$ according to previous measurements in this channel.

\begin{table*}
\renewcommand\arraystretch{1.5}
\caption{Theoretical and experimental parameters for the $s$-wave and $p$-wave FRs in the $^{85}$Rb$|2,-2\rangle+^{87}$Rb$|1,-1\rangle$ and $^{85}$Rb$|2,+2\rangle+^{87}$Rb$|1,+1\rangle$ channels for magnetic field in the range of 0-900\,G.}
\begin{tabular}{p{2.3cm}p{0.7cm}p{1.7cm}p{1.6cm}p{1.3cm}p{1.4cm}p{1.4cm}p{1.4cm}p{1.2cm}}
\hline
\hline
Channel &      $l$ &   $B_{0l}^{\textmd{expt}}$ (G)    &   $\delta_{Bl}^{\textmd{expt}}$(G)${^*}$  &   $B_{0l}$ (G)   &   $\Delta_{Bl}$ (G)    &   $\tilde{a}_{\textmd{bgl}}/\bar{a}_{l} $    &   $\delta\mu_{l}/\mu_{B}$   &   $\zeta_{\textmd{res}}$\\
\hline
$^{85}$Rb$|2, -2\rangle$    &     0   &   264.8(2)    &   2.3    &   262.2   &   -2.6  &   1.984   &   -1.752  &   2.719\\
+$^{87}$Rb$|1, -1\rangle$   &     0   &   372(1)  &   18    &   375.9   &   -26.4  &   2.458   &   -0.9172 &   17.97\\
                            &     0   &   675(2)   &   73    &   666.7   &   -64.7  &   3.375   &   -1.4362   & 94.58   \\
                            &     1   &   257.2(2)    &   2.3    &   258.3   &   26.4   &   -10.36  &   -1.736  &   -85.95\\
                            &     1   &   318.1(2)   &   0.9    &   309.8   &   59.0   &   -2.845  &   -1.817   & -55.23   \\
                            &     1   &   599.3(2)   &   1.7    &   600.7   &   35.2   &   -2.847  &   -0.5112   & -9.252   \\
\hline
$^{85}$Rb$|2, +2\rangle$    &     0   &   569(1)   &   66    &   530.4   &   70.0   &   2.554   &   1.574   & 84.93    \\
+$^{87}$Rb$|1, +1\rangle$   &     0   &   -       &   -       &   701.1   &   0.3  &   1.518   &   1.879   & 0.2966    \\
                            &     1   &   669.0(2)$^{\dag}$   &   0.04    &   668.3   &   2.0   &   1.998   &   1.873  &  -1.360    \\
                            &     1   &   823.3(7)   &   13    &   813.0   &   -142.8  &   -8.824  &   2.453   & -559.0    \\
\hline
\hline
\end{tabular}
\\
\begin{flushleft}
* The parameter $\delta_{Bl}^{\textmd{expt}}$ represents the measured FWHM of a loss feature. It is different from $\Delta_{Bl}$ defined by Eq.~(\ref{scattEqn}).

 \dag~We cannot confirm unambiguously if this FR corresponds correctly to the prediction since the doublet structure is not observed.
\end{flushleft}
\label{table}
\end{table*}

\section{Discussion}
From Table~\ref{table}, we note certain discrepancies between the predictions and the measured results of the FR positions. These discrepancies are generally larger than those offered by numerical coupled-channel calculations facilitated with full knowledge of the relevant molecular potentials~\cite{wille2008exploring,blackley2013feshbach,cho2013feshbach,schuster2012feshbach,repp2013observation,cho2013feshbach, wang2013observation}. They arise from the fact that the MQDT we adopt employs the analytic solutions for an ideal long-range $-1/R^6$ potential, while the contributions of the long-range $-1/R^8$ and $-1/R^{10}$ need to be included  for interatomic separations shorter than $\sim0.3\beta_6$ (about 2.5\,nm for Rb atoms). These discrepancies should be smaller if the energy- and partial-wave-dependent corrections to the quantum defects are considered. One of the advantages of the semi-analytic MQDT is that it does not require full knowledge of the molecular potentials, but only the $C_6$ coefficient, and the singlet and triplet $s$-wave scattering lengths, all of which can be determined with the measurement of a few FR positions. More importantly, the semi-analytic nature of the theory makes it more efficient and reliable for predicting non-s-wave FRs, particularly narrow resonances, since our approach is immune to errors that can plague numerical calculations when handling the classically forbidden regime in the centrifugal barriers.

\section{Conclusion and outlook}

In conclusion, our theory and experiment have combined to verify the existence of broad Feshbach resonances in nonzero partial waves for a $^{85}$Rb and $^{87}$Rb mixture. In particular, our results show that the previously observed $p$-wave resonance at 257.2(2)\,G in the $^{85}$Rb$|2,-2\rangle+^{87}$Rb$|1,-1\rangle$ channel~\cite{papp2006observation} is an entrance-channel dominated broad resonance with $|\zeta_\mathrm{res}|\approx 86$. Within the same channel, we observe two new broad $p$-wave resonances at 318.1(2)\,G and 599.3(2)\,G, with $|\zeta_\mathrm{res}|\approx 55$ and 9, respectively. In the $^{85}$Rb$|2,+2\rangle+^{87}$Rb$|1,+1\rangle$ channel, an even broader $p$-wave resonance at 823.3(7)\,G with $|\zeta_\mathrm{res}|\approx559$ is discovered. The existence of such resonances opens up new possibilities for investigating universal few-body and many-body behaviors in a bosonic atom mixture with strong coupling in nonzero partial waves. As a matter of fact, our trap loss measurement is already a ``measure'' of the three-body recombination. Even without carrying out further experiments to extract the rates explicitly, we can already expect certain qualitative features, such as the asymmetric line shape, to persist. And such a conclusion would already has a physics meaning implying that three-body recombination at ultracold temperatures is dominated by the indirect process (successive pairwise interaction via resonance)~\cite{Pack98,Forrey15} if there exists a $p$ resonance within $k_B T$ above the threshold~\cite{Gaoupb}.

The remaining discrepancies between theory and experiment, as seen in Table~\ref{table}, and the splitting, shown in Fig.~\ref{Pdoublet}, provide motivation for better theories of atomic interactions. The differences between theoretical and experimental resonance positions are due to the energy and the partial wave dependences of the short range QDT parameters that are not included in the present calculation. While it is possible to incorporate such variations by fitting to the experimental data, doing so would require more parameters than the absolute minimum, and would further sacrifice valuable physical information. We believe that all relevant variations are due to the $-C_8/R^8$ potential, and the best way forward is likely through a multichannel and multiscale QDT, with the multiscale aspect being conceptually similar to what we have recently developed for the $-C_1/R-C_4/R^4$ potential~\cite{Fu16}. Such a theory would have one, and only one, more parameter which is $C_8$, and the differences between theory and experiment in Table~\ref{table} can be expected to give us a \textit{measurement} of this physical parameter.

\begin{acknowledgments}
This work is supported at Tsinghua by MOST 973 (Contracts No. 2013CB922004 and No. 2014CB921403) of the National Key Basic Research Program of China, by NSFC (Contracts No. 91121005, No. 91421305, No. 11574177, No. 11374176, No. 11328404, No. 91636213), and at Toledo by the US NSF (through Contracts No. PHY-1306407 and NO. PHY-1607256).
\end{acknowledgments}


\begin{thebibliography}{66}%
\makeatletter
\providecommand \@ifxundefined [1]{%
 \@ifx{#1\undefined}
}%
\providecommand \@ifnum [1]{%
 \ifnum #1\expandafter \@firstoftwo
 \else \expandafter \@secondoftwo
 \fi
}%
\providecommand \@ifx [1]{%
 \ifx #1\expandafter \@firstoftwo
 \else \expandafter \@secondoftwo
 \fi
}%
\providecommand \natexlab [1]{#1}%
\providecommand \enquote  [1]{``#1''}%
\providecommand \bibnamefont  [1]{#1}%
\providecommand \bibfnamefont [1]{#1}%
\providecommand \citenamefont [1]{#1}%
\providecommand \href@noop [0]{\@secondoftwo}%
\providecommand \href [0]{\begingroup \@sanitize@url \@href}%
\providecommand \@href[1]{\@@startlink{#1}\@@href}%
\providecommand \@@href[1]{\endgroup#1\@@endlink}%
\providecommand \@sanitize@url [0]{\catcode `\\12\catcode `\$12\catcode
  `\&12\catcode `\#12\catcode `\^12\catcode `\_12\catcode `\%12\relax}%
\providecommand \@@startlink[1]{}%
\providecommand \@@endlink[0]{}%
\providecommand \url  [0]{\begingroup\@sanitize@url \@url }%
\providecommand \@url [1]{\endgroup\@href {#1}{\urlprefix }}%
\providecommand \urlprefix  [0]{URL }%
\providecommand \Eprint [0]{\href }%
\providecommand \doibase [0]{http://dx.doi.org/}%
\providecommand \selectlanguage [0]{\@gobble}%
\providecommand \bibinfo  [0]{\@secondoftwo}%
\providecommand \bibfield  [0]{\@secondoftwo}%
\providecommand \translation [1]{[#1]}%
\providecommand \BibitemOpen [0]{}%
\providecommand \bibitemStop [0]{}%
\providecommand \bibitemNoStop [0]{.\EOS\space}%
\providecommand \EOS [0]{\spacefactor3000\relax}%
\providecommand \BibitemShut  [1]{\csname bibitem#1\endcsname}%
\let\auto@bib@innerbib\@empty
\bibitem [{\citenamefont {Chin}\ \emph {et~al.}(2010)\citenamefont {Chin},
  \citenamefont {Grimm}, \citenamefont {Julienne},\ and\ \citenamefont
  {Tiesinga}}]{chin2010feshbach}%
  \BibitemOpen
  \bibfield  {author} {\bibinfo {author} {\bibfnamefont {C.}~\bibnamefont
  {Chin}}, \bibinfo {author} {\bibfnamefont {R.}~\bibnamefont {Grimm}},
  \bibinfo {author} {\bibfnamefont {P.}~\bibnamefont {Julienne}}, \ and\
  \bibinfo {author} {\bibfnamefont {E.}~\bibnamefont {Tiesinga}},\ }\href
  {\doibase 10.1103/RevModPhys.82.1225} {\bibfield  {journal} {\bibinfo
  {journal} {Rev. Mod. Phys.}\ }\textbf {\bibinfo {volume} {82}},\ \bibinfo
  {pages} {1225} (\bibinfo {year} {2010})}\BibitemShut {NoStop}%
\bibitem [{\citenamefont {Cornish}\ \emph {et~al.}(2000)\citenamefont
  {Cornish}, \citenamefont {Claussen}, \citenamefont {Roberts}, \citenamefont
  {Cornell},\ and\ \citenamefont {Wieman}}]{Cornish2000TuningInteractions}%
  \BibitemOpen
  \bibfield  {author} {\bibinfo {author} {\bibfnamefont {S.~L.}\ \bibnamefont
  {Cornish}}, \bibinfo {author} {\bibfnamefont {N.~R.}\ \bibnamefont
  {Claussen}}, \bibinfo {author} {\bibfnamefont {J.~L.}\ \bibnamefont
  {Roberts}}, \bibinfo {author} {\bibfnamefont {E.~A.}\ \bibnamefont
  {Cornell}}, \ and\ \bibinfo {author} {\bibfnamefont {C.~E.}\ \bibnamefont
  {Wieman}},\ }\href {\doibase 10.1103/PhysRevLett.85.1795} {\bibfield
  {journal} {\bibinfo  {journal} {Phys. Rev. Lett.}\ }\textbf {\bibinfo
  {volume} {85}},\ \bibinfo {pages} {1795} (\bibinfo {year}
  {2000})}\BibitemShut {NoStop}%
\bibitem [{\citenamefont {Strecker}\ \emph {et~al.}(2002)\citenamefont
  {Strecker}, \citenamefont {Partridge}, \citenamefont {Truscott},\ and\
  \citenamefont {Hulet}}]{Strecker2002Li7}%
  \BibitemOpen
  \bibfield  {author} {\bibinfo {author} {\bibfnamefont {K.~E.}\ \bibnamefont
  {Strecker}}, \bibinfo {author} {\bibfnamefont {G.~B.}\ \bibnamefont
  {Partridge}}, \bibinfo {author} {\bibfnamefont {A.~G.}\ \bibnamefont
  {Truscott}}, \ and\ \bibinfo {author} {\bibfnamefont {R.~G.}\ \bibnamefont
  {Hulet}},\ }\href {\doibase 10.1038/nature747} {\bibfield  {journal}
  {\bibinfo  {journal} {Nature}\ }\textbf {\bibinfo {volume} {417}},\ \bibinfo
  {pages} {150} (\bibinfo {year} {2002})}\BibitemShut {NoStop}%
\bibitem [{\citenamefont {Weber}\ \emph {et~al.}(2003)\citenamefont {Weber},
  \citenamefont {Herbig}, \citenamefont {Mark}, \citenamefont {N{\"a}gerl},\
  and\ \citenamefont {Grimm}}]{Weber2003CsBEC}%
  \BibitemOpen
  \bibfield  {author} {\bibinfo {author} {\bibfnamefont {T.}~\bibnamefont
  {Weber}}, \bibinfo {author} {\bibfnamefont {J.}~\bibnamefont {Herbig}},
  \bibinfo {author} {\bibfnamefont {M.}~\bibnamefont {Mark}}, \bibinfo {author}
  {\bibfnamefont {H.-C.}\ \bibnamefont {N{\"a}gerl}}, \ and\ \bibinfo {author}
  {\bibfnamefont {R.}~\bibnamefont {Grimm}},\ }\href {\doibase
  10.1126/science.1079699} {\bibfield  {journal} {\bibinfo  {journal}
  {Science}\ }\textbf {\bibinfo {volume} {299}},\ \bibinfo {pages} {232}
  (\bibinfo {year} {2003})}\BibitemShut {NoStop}%
\bibitem [{\citenamefont {Kraemer}\ \emph {et~al.}(2006)\citenamefont
  {Kraemer}, \citenamefont {Mark}, \citenamefont {Waldburger}, \citenamefont
  {Danzl}, \citenamefont {Chin}, \citenamefont {Engeser}, \citenamefont
  {Lange}, \citenamefont {Pilch}, \citenamefont {Jaakkola}, \citenamefont
  {N{\"a}gerl},\ and\ \citenamefont {Grimm}}]{Kraemer2006}%
  \BibitemOpen
  \bibfield  {author} {\bibinfo {author} {\bibfnamefont {T.}~\bibnamefont
  {Kraemer}}, \bibinfo {author} {\bibfnamefont {M.}~\bibnamefont {Mark}},
  \bibinfo {author} {\bibfnamefont {P.}~\bibnamefont {Waldburger}}, \bibinfo
  {author} {\bibfnamefont {J.~G.}\ \bibnamefont {Danzl}}, \bibinfo {author}
  {\bibfnamefont {C.}~\bibnamefont {Chin}}, \bibinfo {author} {\bibfnamefont
  {B.}~\bibnamefont {Engeser}}, \bibinfo {author} {\bibfnamefont {A.~D.}\
  \bibnamefont {Lange}}, \bibinfo {author} {\bibfnamefont {K.}~\bibnamefont
  {Pilch}}, \bibinfo {author} {\bibfnamefont {A.}~\bibnamefont {Jaakkola}},
  \bibinfo {author} {\bibfnamefont {H.-C.}\ \bibnamefont {N{\"a}gerl}}, \ and\
  \bibinfo {author} {\bibfnamefont {R.}~\bibnamefont {Grimm}},\ }\href
  {\doibase 10.1038/nature04626} {\bibfield  {journal} {\bibinfo  {journal}
  {Nature}\ }\textbf {\bibinfo {volume} {440}},\ \bibinfo {pages} {315}
  (\bibinfo {year} {2006})}\BibitemShut {NoStop}%
\bibitem [{\citenamefont {von Stecher}\ \emph {et~al.}(2009)\citenamefont {von
  Stecher}, \citenamefont {D'Incao},\ and\ \citenamefont
  {Greene}}]{vonStecher2009}%
  \BibitemOpen
  \bibfield  {author} {\bibinfo {author} {\bibfnamefont {J.}~\bibnamefont {von
  Stecher}}, \bibinfo {author} {\bibfnamefont {J.~P.}\ \bibnamefont {D'Incao}},
  \ and\ \bibinfo {author} {\bibfnamefont {C.~H.}\ \bibnamefont {Greene}},\
  }\href {\doibase 10.1038/nphys1253} {\bibfield  {journal} {\bibinfo
  {journal} {Nat Phys}\ }\textbf {\bibinfo {volume} {5}},\ \bibinfo {pages}
  {417} (\bibinfo {year} {2009})}\BibitemShut {NoStop}%
\bibitem [{\citenamefont {Wenz}\ \emph {et~al.}(2013)\citenamefont {Wenz},
  \citenamefont {Z{\"u}rn}, \citenamefont {Murmann}, \citenamefont {Brouzos},
  \citenamefont {Lompe},\ and\ \citenamefont {Jochim}}]{Jochim2013FewAtoms}%
  \BibitemOpen
  \bibfield  {author} {\bibinfo {author} {\bibfnamefont {A.~N.}\ \bibnamefont
  {Wenz}}, \bibinfo {author} {\bibfnamefont {G.}~\bibnamefont {Z{\"u}rn}},
  \bibinfo {author} {\bibfnamefont {S.}~\bibnamefont {Murmann}}, \bibinfo
  {author} {\bibfnamefont {I.}~\bibnamefont {Brouzos}}, \bibinfo {author}
  {\bibfnamefont {T.}~\bibnamefont {Lompe}}, \ and\ \bibinfo {author}
  {\bibfnamefont {S.}~\bibnamefont {Jochim}},\ }\href {\doibase
  10.1126/science.1240516} {\bibfield  {journal} {\bibinfo  {journal}
  {Science}\ }\textbf {\bibinfo {volume} {342}},\ \bibinfo {pages} {457}
  (\bibinfo {year} {2013})}\BibitemShut {NoStop}%
\bibitem [{\citenamefont {Regal}\ \emph {et~al.}(2003)\citenamefont {Regal},
  \citenamefont {Ticknor}, \citenamefont {Bohn},\ and\ \citenamefont
  {Jin}}]{regal2003creation}%
  \BibitemOpen
  \bibfield  {author} {\bibinfo {author} {\bibfnamefont {C.~A.}\ \bibnamefont
  {Regal}}, \bibinfo {author} {\bibfnamefont {C.}~\bibnamefont {Ticknor}},
  \bibinfo {author} {\bibfnamefont {J.~L.}\ \bibnamefont {Bohn}}, \ and\
  \bibinfo {author} {\bibfnamefont {D.~S.}\ \bibnamefont {Jin}},\ }\href
  {\doibase 10.1038/nature01738} {\bibfield  {journal} {\bibinfo  {journal}
  {Nature}\ }\textbf {\bibinfo {volume} {424}},\ \bibinfo {pages} {47}
  (\bibinfo {year} {2003})}\BibitemShut {NoStop}%
\bibitem [{\citenamefont {Ni}\ \emph {et~al.}(2008)\citenamefont {Ni},
  \citenamefont {Ospelkaus}, \citenamefont {de~Miranda}, \citenamefont
  {Pe{\textquoteright}er}, \citenamefont {Neyenhuis}, \citenamefont {Zirbel},
  \citenamefont {Kotochigova}, \citenamefont {Julienne}, \citenamefont {Jin},\
  and\ \citenamefont {Ye}}]{ni2008high}%
  \BibitemOpen
  \bibfield  {author} {\bibinfo {author} {\bibfnamefont {K.-K.}\ \bibnamefont
  {Ni}}, \bibinfo {author} {\bibfnamefont {S.}~\bibnamefont {Ospelkaus}},
  \bibinfo {author} {\bibfnamefont {M.~H.~G.}\ \bibnamefont {de~Miranda}},
  \bibinfo {author} {\bibfnamefont {A.}~\bibnamefont {Pe{\textquoteright}er}},
  \bibinfo {author} {\bibfnamefont {B.}~\bibnamefont {Neyenhuis}}, \bibinfo
  {author} {\bibfnamefont {J.~J.}\ \bibnamefont {Zirbel}}, \bibinfo {author}
  {\bibfnamefont {S.}~\bibnamefont {Kotochigova}}, \bibinfo {author}
  {\bibfnamefont {P.~S.}\ \bibnamefont {Julienne}}, \bibinfo {author}
  {\bibfnamefont {D.~S.}\ \bibnamefont {Jin}}, \ and\ \bibinfo {author}
  {\bibfnamefont {J.}~\bibnamefont {Ye}},\ }\href {\doibase
  10.1126/science.1163861} {\bibfield  {journal} {\bibinfo  {journal}
  {Science}\ }\textbf {\bibinfo {volume} {322}},\ \bibinfo {pages} {231}
  (\bibinfo {year} {2008})}\BibitemShut {NoStop}%
\bibitem [{\citenamefont {Lang}\ \emph {et~al.}(2008)\citenamefont {Lang},
  \citenamefont {Winkler}, \citenamefont {Strauss}, \citenamefont {Grimm},\
  and\ \citenamefont {Denschlag}}]{lang2008ultracold}%
  \BibitemOpen
  \bibfield  {author} {\bibinfo {author} {\bibfnamefont {F.}~\bibnamefont
  {Lang}}, \bibinfo {author} {\bibfnamefont {K.}~\bibnamefont {Winkler}},
  \bibinfo {author} {\bibfnamefont {C.}~\bibnamefont {Strauss}}, \bibinfo
  {author} {\bibfnamefont {R.}~\bibnamefont {Grimm}}, \ and\ \bibinfo {author}
  {\bibfnamefont {J.~H.}\ \bibnamefont {Denschlag}},\ }\href {\doibase
  10.1103/PhysRevLett.101.133005} {\bibfield  {journal} {\bibinfo  {journal}
  {Phys. Rev. Lett.}\ }\textbf {\bibinfo {volume} {101}},\ \bibinfo {pages}
  {133005} (\bibinfo {year} {2008})}\BibitemShut {NoStop}%
\bibitem [{\citenamefont {Danzl}\ \emph {et~al.}(2010)\citenamefont {Danzl},
  \citenamefont {Mark}, \citenamefont {Haller}, \citenamefont {Gustavsson},
  \citenamefont {Hart}, \citenamefont {Aldegunde}, \citenamefont {Hutson},\
  and\ \citenamefont {N{\"a}gerl}}]{danzl2010ultracold}%
  \BibitemOpen
  \bibfield  {author} {\bibinfo {author} {\bibfnamefont {J.~G.}\ \bibnamefont
  {Danzl}}, \bibinfo {author} {\bibfnamefont {M.~J.}\ \bibnamefont {Mark}},
  \bibinfo {author} {\bibfnamefont {E.}~\bibnamefont {Haller}}, \bibinfo
  {author} {\bibfnamefont {M.}~\bibnamefont {Gustavsson}}, \bibinfo {author}
  {\bibfnamefont {R.}~\bibnamefont {Hart}}, \bibinfo {author} {\bibfnamefont
  {J.}~\bibnamefont {Aldegunde}}, \bibinfo {author} {\bibfnamefont {J.~M.}\
  \bibnamefont {Hutson}}, \ and\ \bibinfo {author} {\bibfnamefont {H.-C.}\
  \bibnamefont {N{\"a}gerl}},\ }\href {\doibase 10.1038/nphys1533} {\bibfield
  {journal} {\bibinfo  {journal} {Nat Phys}\ }\textbf {\bibinfo {volume} {6}},\
  \bibinfo {pages} {265} (\bibinfo {year} {2010})}\BibitemShut {NoStop}%
\bibitem [{\citenamefont {Kinast}\ \emph {et~al.}(2005)\citenamefont {Kinast},
  \citenamefont {Turlapov}, \citenamefont {Thomas}, \citenamefont {Chen},
  \citenamefont {Stajic},\ and\ \citenamefont {Levin}}]{Kinast1296}%
  \BibitemOpen
  \bibfield  {author} {\bibinfo {author} {\bibfnamefont {J.}~\bibnamefont
  {Kinast}}, \bibinfo {author} {\bibfnamefont {A.}~\bibnamefont {Turlapov}},
  \bibinfo {author} {\bibfnamefont {J.~E.}\ \bibnamefont {Thomas}}, \bibinfo
  {author} {\bibfnamefont {Q.}~\bibnamefont {Chen}}, \bibinfo {author}
  {\bibfnamefont {J.}~\bibnamefont {Stajic}}, \ and\ \bibinfo {author}
  {\bibfnamefont {K.}~\bibnamefont {Levin}},\ }\href {\doibase
  10.1126/science.1109220} {\bibfield  {journal} {\bibinfo  {journal}
  {Science}\ }\textbf {\bibinfo {volume} {307}},\ \bibinfo {pages} {1296}
  (\bibinfo {year} {2005})}\BibitemShut {NoStop}%
\bibitem [{\citenamefont {Horikoshi}\ \emph {et~al.}(2010)\citenamefont
  {Horikoshi}, \citenamefont {Nakajima}, \citenamefont {Ueda},\ and\
  \citenamefont {Mukaiyama}}]{Horikoshi442}%
  \BibitemOpen
  \bibfield  {author} {\bibinfo {author} {\bibfnamefont {M.}~\bibnamefont
  {Horikoshi}}, \bibinfo {author} {\bibfnamefont {S.}~\bibnamefont {Nakajima}},
  \bibinfo {author} {\bibfnamefont {M.}~\bibnamefont {Ueda}}, \ and\ \bibinfo
  {author} {\bibfnamefont {T.}~\bibnamefont {Mukaiyama}},\ }\href {\doibase
  10.1126/science.1183012} {\bibfield  {journal} {\bibinfo  {journal}
  {Science}\ }\textbf {\bibinfo {volume} {327}},\ \bibinfo {pages} {442}
  (\bibinfo {year} {2010})}\BibitemShut {NoStop}%
\bibitem [{\citenamefont {Nascimb\`ene}\ \emph {et~al.}(2010)\citenamefont
  {Nascimb\`ene}, \citenamefont {Navon}, \citenamefont {Jiang}, \citenamefont
  {Chevy},\ and\ \citenamefont {Salomon}}]{Nascimbene2010}%
  \BibitemOpen
  \bibfield  {author} {\bibinfo {author} {\bibfnamefont {S.}~\bibnamefont
  {Nascimb\`ene}}, \bibinfo {author} {\bibfnamefont {N.}~\bibnamefont {Navon}},
  \bibinfo {author} {\bibfnamefont {K.~J.}\ \bibnamefont {Jiang}}, \bibinfo
  {author} {\bibfnamefont {F.}~\bibnamefont {Chevy}}, \ and\ \bibinfo {author}
  {\bibfnamefont {C.}~\bibnamefont {Salomon}},\ }\href {\doibase
  10.1038/nature08814} {\bibfield  {journal} {\bibinfo  {journal} {Nature}\
  }\textbf {\bibinfo {volume} {463}},\ \bibinfo {pages} {1057} (\bibinfo {year}
  {2010})}\BibitemShut {NoStop}%
\bibitem [{\citenamefont {Ku}\ \emph {et~al.}(2012)\citenamefont {Ku},
  \citenamefont {Sommer}, \citenamefont {Cheuk},\ and\ \citenamefont
  {Zwierlein}}]{Ku2012EOS}%
  \BibitemOpen
  \bibfield  {author} {\bibinfo {author} {\bibfnamefont {M.~J.~H.}\
  \bibnamefont {Ku}}, \bibinfo {author} {\bibfnamefont {A.~T.}\ \bibnamefont
  {Sommer}}, \bibinfo {author} {\bibfnamefont {L.~W.}\ \bibnamefont {Cheuk}}, \
  and\ \bibinfo {author} {\bibfnamefont {M.~W.}\ \bibnamefont {Zwierlein}},\
  }\href {\doibase 10.1126/science.1214987} {\bibfield  {journal} {\bibinfo
  {journal} {Science}\ }\textbf {\bibinfo {volume} {335}},\ \bibinfo {pages}
  {563} (\bibinfo {year} {2012})}\BibitemShut {NoStop}%
\bibitem [{\citenamefont {Sidorenkov}\ \emph {et~al.}(2013)\citenamefont
  {Sidorenkov}, \citenamefont {Tey}, \citenamefont {Grimm}, \citenamefont
  {Hou}, \citenamefont {Pitaevskii},\ and\ \citenamefont
  {Stringari}}]{Sidorenkov2013}%
  \BibitemOpen
  \bibfield  {author} {\bibinfo {author} {\bibfnamefont {L.~A.}\ \bibnamefont
  {Sidorenkov}}, \bibinfo {author} {\bibfnamefont {M.~K.}\ \bibnamefont {Tey}},
  \bibinfo {author} {\bibfnamefont {R.}~\bibnamefont {Grimm}}, \bibinfo
  {author} {\bibfnamefont {Y.-H.}\ \bibnamefont {Hou}}, \bibinfo {author}
  {\bibfnamefont {L.}~\bibnamefont {Pitaevskii}}, \ and\ \bibinfo {author}
  {\bibfnamefont {S.}~\bibnamefont {Stringari}},\ }\href
  {http://dx.doi.org/10.1038/nature12136} {\bibfield  {journal} {\bibinfo
  {journal} {Nature}\ }\textbf {\bibinfo {volume} {498}},\ \bibinfo {pages}
  {78} (\bibinfo {year} {2013})},\ \bibinfo {note} {letter}\BibitemShut
  {NoStop}%
\bibitem [{\citenamefont {Tey}\ \emph {et~al.}(2013)\citenamefont {Tey},
  \citenamefont {Sidorenkov}, \citenamefont {Guajardo}, \citenamefont {Grimm},
  \citenamefont {Ku}, \citenamefont {Zwierlein}, \citenamefont {Hou},
  \citenamefont {Pitaevskii},\ and\ \citenamefont
  {Stringari}}]{Tey2013CollectiveModes}%
  \BibitemOpen
  \bibfield  {author} {\bibinfo {author} {\bibfnamefont {M.~K.}\ \bibnamefont
  {Tey}}, \bibinfo {author} {\bibfnamefont {L.~A.}\ \bibnamefont {Sidorenkov}},
  \bibinfo {author} {\bibfnamefont {E.~R.~S.}\ \bibnamefont {Guajardo}},
  \bibinfo {author} {\bibfnamefont {R.}~\bibnamefont {Grimm}}, \bibinfo
  {author} {\bibfnamefont {M.~J.~H.}\ \bibnamefont {Ku}}, \bibinfo {author}
  {\bibfnamefont {M.~W.}\ \bibnamefont {Zwierlein}}, \bibinfo {author}
  {\bibfnamefont {Y.-H.}\ \bibnamefont {Hou}}, \bibinfo {author} {\bibfnamefont
  {L.}~\bibnamefont {Pitaevskii}}, \ and\ \bibinfo {author} {\bibfnamefont
  {S.}~\bibnamefont {Stringari}},\ }\href {\doibase
  10.1103/PhysRevLett.110.055303} {\bibfield  {journal} {\bibinfo  {journal}
  {Phys. Rev. Lett.}\ }\textbf {\bibinfo {volume} {110}},\ \bibinfo {pages}
  {055303} (\bibinfo {year} {2013})}\BibitemShut {NoStop}%
\bibitem [{\citenamefont {Guajardo}\ \emph {et~al.}(2013)\citenamefont
  {Guajardo}, \citenamefont {Tey}, \citenamefont {Sidorenkov},\ and\
  \citenamefont {Grimm}}]{Edmundo2013HigherOrder}%
  \BibitemOpen
  \bibfield  {author} {\bibinfo {author} {\bibfnamefont {E.~R.~S.}\
  \bibnamefont {Guajardo}}, \bibinfo {author} {\bibfnamefont {M.~K.}\
  \bibnamefont {Tey}}, \bibinfo {author} {\bibfnamefont {L.~A.}\ \bibnamefont
  {Sidorenkov}}, \ and\ \bibinfo {author} {\bibfnamefont {R.}~\bibnamefont
  {Grimm}},\ }\href {\doibase 10.1103/PhysRevA.87.063601} {\bibfield  {journal}
  {\bibinfo  {journal} {Phys. Rev. A}\ }\textbf {\bibinfo {volume} {87}},\
  \bibinfo {pages} {063601} (\bibinfo {year} {2013})}\BibitemShut {NoStop}%
\bibitem [{\citenamefont {Chin}\ \emph {et~al.}(2000)\citenamefont {Chin},
  \citenamefont {Vuleti\ifmmode~\acute{c}\else \'{c}\fi{}}, \citenamefont
  {Kerman},\ and\ \citenamefont {Chu}}]{chin2000high}%
  \BibitemOpen
  \bibfield  {author} {\bibinfo {author} {\bibfnamefont {C.}~\bibnamefont
  {Chin}}, \bibinfo {author} {\bibfnamefont {V.}~\bibnamefont
  {Vuleti\ifmmode~\acute{c}\else \'{c}\fi{}}}, \bibinfo {author} {\bibfnamefont
  {A.~J.}\ \bibnamefont {Kerman}}, \ and\ \bibinfo {author} {\bibfnamefont
  {S.}~\bibnamefont {Chu}},\ }\href {\doibase 10.1103/PhysRevLett.85.2717}
  {\bibfield  {journal} {\bibinfo  {journal} {Phys. Rev. Lett.}\ }\textbf
  {\bibinfo {volume} {85}},\ \bibinfo {pages} {2717} (\bibinfo {year}
  {2000})}\BibitemShut {NoStop}%
\bibitem [{\citenamefont {Dieckmann}\ \emph {et~al.}(2002)\citenamefont
  {Dieckmann}, \citenamefont {Stan}, \citenamefont {Gupta}, \citenamefont
  {Hadzibabic}, \citenamefont {Schunck},\ and\ \citenamefont
  {Ketterle}}]{PhysRevLett.89.203201}%
  \BibitemOpen
  \bibfield  {author} {\bibinfo {author} {\bibfnamefont {K.}~\bibnamefont
  {Dieckmann}}, \bibinfo {author} {\bibfnamefont {C.~A.}\ \bibnamefont {Stan}},
  \bibinfo {author} {\bibfnamefont {S.}~\bibnamefont {Gupta}}, \bibinfo
  {author} {\bibfnamefont {Z.}~\bibnamefont {Hadzibabic}}, \bibinfo {author}
  {\bibfnamefont {C.~H.}\ \bibnamefont {Schunck}}, \ and\ \bibinfo {author}
  {\bibfnamefont {W.}~\bibnamefont {Ketterle}},\ }\href {\doibase
  10.1103/PhysRevLett.89.203201} {\bibfield  {journal} {\bibinfo  {journal}
  {Phys. Rev. Lett.}\ }\textbf {\bibinfo {volume} {89}},\ \bibinfo {pages}
  {203201} (\bibinfo {year} {2002})}\BibitemShut {NoStop}%
\bibitem [{\citenamefont {Loftus}\ \emph {et~al.}(2002)\citenamefont {Loftus},
  \citenamefont {Regal}, \citenamefont {Ticknor}, \citenamefont {Bohn},\ and\
  \citenamefont {Jin}}]{PhysRevLett.88.173201}%
  \BibitemOpen
  \bibfield  {author} {\bibinfo {author} {\bibfnamefont {T.}~\bibnamefont
  {Loftus}}, \bibinfo {author} {\bibfnamefont {C.~A.}\ \bibnamefont {Regal}},
  \bibinfo {author} {\bibfnamefont {C.}~\bibnamefont {Ticknor}}, \bibinfo
  {author} {\bibfnamefont {J.~L.}\ \bibnamefont {Bohn}}, \ and\ \bibinfo
  {author} {\bibfnamefont {D.~S.}\ \bibnamefont {Jin}},\ }\href {\doibase
  10.1103/PhysRevLett.88.173201} {\bibfield  {journal} {\bibinfo  {journal}
  {Phys. Rev. Lett.}\ }\textbf {\bibinfo {volume} {88}},\ \bibinfo {pages}
  {173201} (\bibinfo {year} {2002})}\BibitemShut {NoStop}%
\bibitem [{\citenamefont {Knoop}\ \emph {et~al.}(2011)\citenamefont {Knoop},
  \citenamefont {Schuster}, \citenamefont {Scelle}, \citenamefont {Trautmann},
  \citenamefont {Appmeier}, \citenamefont {Oberthaler}, \citenamefont
  {Tiesinga},\ and\ \citenamefont {Tiemann}}]{knoop2011feshbach}%
  \BibitemOpen
  \bibfield  {author} {\bibinfo {author} {\bibfnamefont {S.}~\bibnamefont
  {Knoop}}, \bibinfo {author} {\bibfnamefont {T.}~\bibnamefont {Schuster}},
  \bibinfo {author} {\bibfnamefont {R.}~\bibnamefont {Scelle}}, \bibinfo
  {author} {\bibfnamefont {A.}~\bibnamefont {Trautmann}}, \bibinfo {author}
  {\bibfnamefont {J.}~\bibnamefont {Appmeier}}, \bibinfo {author}
  {\bibfnamefont {M.~K.}\ \bibnamefont {Oberthaler}}, \bibinfo {author}
  {\bibfnamefont {E.}~\bibnamefont {Tiesinga}}, \ and\ \bibinfo {author}
  {\bibfnamefont {E.}~\bibnamefont {Tiemann}},\ }\href {\doibase
  10.1103/PhysRevA.83.042704} {\bibfield  {journal} {\bibinfo  {journal} {Phys.
  Rev. A}\ }\textbf {\bibinfo {volume} {83}},\ \bibinfo {pages} {042704}
  (\bibinfo {year} {2011})}\BibitemShut {NoStop}%
\bibitem [{\citenamefont {Blackley}\ \emph {et~al.}(2013)\citenamefont
  {Blackley}, \citenamefont {Le~Sueur}, \citenamefont {Hutson}, \citenamefont
  {McCarron}, \citenamefont {K\"oppinger}, \citenamefont {Cho}, \citenamefont
  {Jenkin},\ and\ \citenamefont {Cornish}}]{blackley2013feshbach}%
  \BibitemOpen
  \bibfield  {author} {\bibinfo {author} {\bibfnamefont {C.~L.}\ \bibnamefont
  {Blackley}}, \bibinfo {author} {\bibfnamefont {C.~R.}\ \bibnamefont
  {Le~Sueur}}, \bibinfo {author} {\bibfnamefont {J.~M.}\ \bibnamefont
  {Hutson}}, \bibinfo {author} {\bibfnamefont {D.~J.}\ \bibnamefont
  {McCarron}}, \bibinfo {author} {\bibfnamefont {M.~P.}\ \bibnamefont
  {K\"oppinger}}, \bibinfo {author} {\bibfnamefont {H.-W.}\ \bibnamefont
  {Cho}}, \bibinfo {author} {\bibfnamefont {D.~L.}\ \bibnamefont {Jenkin}}, \
  and\ \bibinfo {author} {\bibfnamefont {S.~L.}\ \bibnamefont {Cornish}},\
  }\href {\doibase 10.1103/PhysRevA.87.033611} {\bibfield  {journal} {\bibinfo
  {journal} {Phys. Rev. A}\ }\textbf {\bibinfo {volume} {87}},\ \bibinfo
  {pages} {033611} (\bibinfo {year} {2013})}\BibitemShut {NoStop}%
\bibitem [{\citenamefont {Papp}\ and\ \citenamefont
  {Wieman}(2006)}]{papp2006observation}%
  \BibitemOpen
  \bibfield  {author} {\bibinfo {author} {\bibfnamefont {S.~B.}\ \bibnamefont
  {Papp}}\ and\ \bibinfo {author} {\bibfnamefont {C.~E.}\ \bibnamefont
  {Wieman}},\ }\href {\doibase 10.1103/PhysRevLett.97.180404} {\bibfield
  {journal} {\bibinfo  {journal} {Phys. Rev. Lett.}\ }\textbf {\bibinfo
  {volume} {97}},\ \bibinfo {pages} {180404} (\bibinfo {year}
  {2006})}\BibitemShut {NoStop}%
\bibitem [{\citenamefont {Ferlaino}\ \emph {et~al.}(2006)\citenamefont
  {Ferlaino}, \citenamefont {D'Errico}, \citenamefont {Roati}, \citenamefont
  {Zaccanti}, \citenamefont {Inguscio}, \citenamefont {Modugno},\ and\
  \citenamefont {Simoni}}]{ferlaino2006feshbach}%
  \BibitemOpen
  \bibfield  {author} {\bibinfo {author} {\bibfnamefont {F.}~\bibnamefont
  {Ferlaino}}, \bibinfo {author} {\bibfnamefont {C.}~\bibnamefont {D'Errico}},
  \bibinfo {author} {\bibfnamefont {G.}~\bibnamefont {Roati}}, \bibinfo
  {author} {\bibfnamefont {M.}~\bibnamefont {Zaccanti}}, \bibinfo {author}
  {\bibfnamefont {M.}~\bibnamefont {Inguscio}}, \bibinfo {author}
  {\bibfnamefont {G.}~\bibnamefont {Modugno}}, \ and\ \bibinfo {author}
  {\bibfnamefont {A.}~\bibnamefont {Simoni}},\ }\href {\doibase
  10.1103/PhysRevA.73.040702} {\bibfield  {journal} {\bibinfo  {journal} {Phys.
  Rev. A}\ }\textbf {\bibinfo {volume} {73}},\ \bibinfo {pages} {040702}
  (\bibinfo {year} {2006})}\BibitemShut {NoStop}%
\bibitem [{\citenamefont {Wille}\ \emph {et~al.}(2008)\citenamefont {Wille},
  \citenamefont {Spiegelhalder}, \citenamefont {Kerner}, \citenamefont {Naik},
  \citenamefont {Trenkwalder}, \citenamefont {Hendl}, \citenamefont {Schreck},
  \citenamefont {Grimm}, \citenamefont {Tiecke}, \citenamefont {Walraven},
  \citenamefont {Kokkelmans}, \citenamefont {Tiesinga},\ and\ \citenamefont
  {Julienne}}]{wille2008exploring}%
  \BibitemOpen
  \bibfield  {author} {\bibinfo {author} {\bibfnamefont {E.}~\bibnamefont
  {Wille}}, \bibinfo {author} {\bibfnamefont {F.~M.}\ \bibnamefont
  {Spiegelhalder}}, \bibinfo {author} {\bibfnamefont {G.}~\bibnamefont
  {Kerner}}, \bibinfo {author} {\bibfnamefont {D.}~\bibnamefont {Naik}},
  \bibinfo {author} {\bibfnamefont {A.}~\bibnamefont {Trenkwalder}}, \bibinfo
  {author} {\bibfnamefont {G.}~\bibnamefont {Hendl}}, \bibinfo {author}
  {\bibfnamefont {F.}~\bibnamefont {Schreck}}, \bibinfo {author} {\bibfnamefont
  {R.}~\bibnamefont {Grimm}}, \bibinfo {author} {\bibfnamefont {T.~G.}\
  \bibnamefont {Tiecke}}, \bibinfo {author} {\bibfnamefont {J.~T.~M.}\
  \bibnamefont {Walraven}}, \bibinfo {author} {\bibfnamefont {S.~J. J. M.~F.}\
  \bibnamefont {Kokkelmans}}, \bibinfo {author} {\bibfnamefont
  {E.}~\bibnamefont {Tiesinga}}, \ and\ \bibinfo {author} {\bibfnamefont
  {P.~S.}\ \bibnamefont {Julienne}},\ }\href {\doibase
  10.1103/PhysRevLett.100.053201} {\bibfield  {journal} {\bibinfo  {journal}
  {Phys. Rev. Lett.}\ }\textbf {\bibinfo {volume} {100}},\ \bibinfo {pages}
  {053201} (\bibinfo {year} {2008})}\BibitemShut {NoStop}%
\bibitem [{\citenamefont {Deh}\ \emph {et~al.}(2008)\citenamefont {Deh},
  \citenamefont {Marzok}, \citenamefont {Zimmermann},\ and\ \citenamefont
  {Courteille}}]{deh2008feshbach}%
  \BibitemOpen
  \bibfield  {author} {\bibinfo {author} {\bibfnamefont {B.}~\bibnamefont
  {Deh}}, \bibinfo {author} {\bibfnamefont {C.}~\bibnamefont {Marzok}},
  \bibinfo {author} {\bibfnamefont {C.}~\bibnamefont {Zimmermann}}, \ and\
  \bibinfo {author} {\bibfnamefont {P.~W.}\ \bibnamefont {Courteille}},\ }\href
  {\doibase 10.1103/PhysRevA.77.010701} {\bibfield  {journal} {\bibinfo
  {journal} {Phys. Rev. A}\ }\textbf {\bibinfo {volume} {77}},\ \bibinfo
  {pages} {010701} (\bibinfo {year} {2008})}\BibitemShut {NoStop}%
\bibitem [{\citenamefont {Repp}\ \emph {et~al.}(2013)\citenamefont {Repp},
  \citenamefont {Pires}, \citenamefont {Ulmanis}, \citenamefont {Heck},
  \citenamefont {Kuhnle}, \citenamefont {Weidem\"uller},\ and\ \citenamefont
  {Tiemann}}]{repp2013observation}%
  \BibitemOpen
  \bibfield  {author} {\bibinfo {author} {\bibfnamefont {M.}~\bibnamefont
  {Repp}}, \bibinfo {author} {\bibfnamefont {R.}~\bibnamefont {Pires}},
  \bibinfo {author} {\bibfnamefont {J.}~\bibnamefont {Ulmanis}}, \bibinfo
  {author} {\bibfnamefont {R.}~\bibnamefont {Heck}}, \bibinfo {author}
  {\bibfnamefont {E.~D.}\ \bibnamefont {Kuhnle}}, \bibinfo {author}
  {\bibfnamefont {M.}~\bibnamefont {Weidem\"uller}}, \ and\ \bibinfo {author}
  {\bibfnamefont {E.}~\bibnamefont {Tiemann}},\ }\href {\doibase
  10.1103/PhysRevA.87.010701} {\bibfield  {journal} {\bibinfo  {journal} {Phys.
  Rev. A}\ }\textbf {\bibinfo {volume} {87}},\ \bibinfo {pages} {010701}
  (\bibinfo {year} {2013})}\BibitemShut {NoStop}%
\bibitem [{\citenamefont {Cho}\ \emph {et~al.}(2013)\citenamefont {Cho},
  \citenamefont {McCarron}, \citenamefont {K\"oppinger}, \citenamefont
  {Jenkin}, \citenamefont {Butler}, \citenamefont {Julienne}, \citenamefont
  {Blackley}, \citenamefont {Le~Sueur}, \citenamefont {Hutson},\ and\
  \citenamefont {Cornish}}]{cho2013feshbach}%
  \BibitemOpen
  \bibfield  {author} {\bibinfo {author} {\bibfnamefont {H.-W.}\ \bibnamefont
  {Cho}}, \bibinfo {author} {\bibfnamefont {D.~J.}\ \bibnamefont {McCarron}},
  \bibinfo {author} {\bibfnamefont {M.~P.}\ \bibnamefont {K\"oppinger}},
  \bibinfo {author} {\bibfnamefont {D.~L.}\ \bibnamefont {Jenkin}}, \bibinfo
  {author} {\bibfnamefont {K.~L.}\ \bibnamefont {Butler}}, \bibinfo {author}
  {\bibfnamefont {P.~S.}\ \bibnamefont {Julienne}}, \bibinfo {author}
  {\bibfnamefont {C.~L.}\ \bibnamefont {Blackley}}, \bibinfo {author}
  {\bibfnamefont {C.~R.}\ \bibnamefont {Le~Sueur}}, \bibinfo {author}
  {\bibfnamefont {J.~M.}\ \bibnamefont {Hutson}}, \ and\ \bibinfo {author}
  {\bibfnamefont {S.~L.}\ \bibnamefont {Cornish}},\ }\href {\doibase
  10.1103/PhysRevA.87.010703} {\bibfield  {journal} {\bibinfo  {journal} {Phys.
  Rev. A}\ }\textbf {\bibinfo {volume} {87}},\ \bibinfo {pages} {010703}
  (\bibinfo {year} {2013})}\BibitemShut {NoStop}%
\bibitem [{\citenamefont {Wang}\ \emph {et~al.}(2013)\citenamefont {Wang},
  \citenamefont {Xiong}, \citenamefont {Li}, \citenamefont {Wang},\ and\
  \citenamefont {Tiemann}}]{wang2013observation}%
  \BibitemOpen
  \bibfield  {author} {\bibinfo {author} {\bibfnamefont {F.}~\bibnamefont
  {Wang}}, \bibinfo {author} {\bibfnamefont {D.}~\bibnamefont {Xiong}},
  \bibinfo {author} {\bibfnamefont {X.}~\bibnamefont {Li}}, \bibinfo {author}
  {\bibfnamefont {D.}~\bibnamefont {Wang}}, \ and\ \bibinfo {author}
  {\bibfnamefont {E.}~\bibnamefont {Tiemann}},\ }\href {\doibase
  10.1103/PhysRevA.87.050702} {\bibfield  {journal} {\bibinfo  {journal} {Phys.
  Rev. A}\ }\textbf {\bibinfo {volume} {87}},\ \bibinfo {pages} {050702}
  (\bibinfo {year} {2013})}\BibitemShut {NoStop}%
\bibitem [{\citenamefont {Gao}\ \emph {et~al.}(2005{\natexlab{a}})\citenamefont
  {Gao}, \citenamefont {Tiesinga}, \citenamefont {Williams},\ and\
  \citenamefont {Julienne}}]{Gao2005MQDT}%
  \BibitemOpen
  \bibfield  {author} {\bibinfo {author} {\bibfnamefont {B.}~\bibnamefont
  {Gao}}, \bibinfo {author} {\bibfnamefont {E.}~\bibnamefont {Tiesinga}},
  \bibinfo {author} {\bibfnamefont {C.~J.}\ \bibnamefont {Williams}}, \ and\
  \bibinfo {author} {\bibfnamefont {P.~S.}\ \bibnamefont {Julienne}},\ }\href
  {\doibase 10.1103/PhysRevA.72.042719} {\bibfield  {journal} {\bibinfo
  {journal} {Phys. Rev. A}\ }\textbf {\bibinfo {volume} {72}},\ \bibinfo
  {pages} {042719} (\bibinfo {year} {2005}{\natexlab{a}})}\BibitemShut
  {NoStop}%
\bibitem [{\citenamefont {Gao}(2011)}]{gao2011analytic}%
  \BibitemOpen
  \bibfield  {author} {\bibinfo {author} {\bibfnamefont {B.}~\bibnamefont
  {Gao}},\ }\href {\doibase 10.1103/PhysRevA.84.022706} {\bibfield  {journal}
  {\bibinfo  {journal} {Phys. Rev. A}\ }\textbf {\bibinfo {volume} {84}},\
  \bibinfo {pages} {022706} (\bibinfo {year} {2011})}\BibitemShut {NoStop}%
\bibitem [{\citenamefont {Makrides}\ and\ \citenamefont
  {Gao}(2014)}]{makrides2014multichannel}%
  \BibitemOpen
  \bibfield  {author} {\bibinfo {author} {\bibfnamefont {C.}~\bibnamefont
  {Makrides}}\ and\ \bibinfo {author} {\bibfnamefont {B.}~\bibnamefont {Gao}},\
  }\href {\doibase 10.1103/PhysRevA.89.062718} {\bibfield  {journal} {\bibinfo
  {journal} {Phys. Rev. A}\ }\textbf {\bibinfo {volume} {89}},\ \bibinfo
  {pages} {062718} (\bibinfo {year} {2014})}\BibitemShut {NoStop}%
\bibitem [{\citenamefont {Suno}\ \emph {et~al.}(2003)\citenamefont {Suno},
  \citenamefont {Esry},\ and\ \citenamefont {Greene}}]{PhysRevLett.90.053202}%
  \BibitemOpen
  \bibfield  {author} {\bibinfo {author} {\bibfnamefont {H.}~\bibnamefont
  {Suno}}, \bibinfo {author} {\bibfnamefont {B.~D.}\ \bibnamefont {Esry}}, \
  and\ \bibinfo {author} {\bibfnamefont {C.~H.}\ \bibnamefont {Greene}},\
  }\href {\doibase 10.1103/PhysRevLett.90.053202} {\bibfield  {journal}
  {\bibinfo  {journal} {Phys. Rev. Lett.}\ }\textbf {\bibinfo {volume} {90}},\
  \bibinfo {pages} {053202} (\bibinfo {year} {2003})}\BibitemShut {NoStop}%
\bibitem [{\citenamefont {G\"unter}\ \emph {et~al.}(2005)\citenamefont
  {G\"unter}, \citenamefont {St\"oferle}, \citenamefont {Moritz}, \citenamefont
  {K\"ohl},\ and\ \citenamefont {Esslinger}}]{PhysRevLett.95.230401}%
  \BibitemOpen
  \bibfield  {author} {\bibinfo {author} {\bibfnamefont {K.}~\bibnamefont
  {G\"unter}}, \bibinfo {author} {\bibfnamefont {T.}~\bibnamefont
  {St\"oferle}}, \bibinfo {author} {\bibfnamefont {H.}~\bibnamefont {Moritz}},
  \bibinfo {author} {\bibfnamefont {M.}~\bibnamefont {K\"ohl}}, \ and\ \bibinfo
  {author} {\bibfnamefont {T.}~\bibnamefont {Esslinger}},\ }\href {\doibase
  10.1103/PhysRevLett.95.230401} {\bibfield  {journal} {\bibinfo  {journal}
  {Phys. Rev. Lett.}\ }\textbf {\bibinfo {volume} {95}},\ \bibinfo {pages}
  {230401} (\bibinfo {year} {2005})}\BibitemShut {NoStop}%
\bibitem [{\citenamefont {Gaebler}\ \emph {et~al.}(2007)\citenamefont
  {Gaebler}, \citenamefont {Stewart}, \citenamefont {Bohn},\ and\ \citenamefont
  {Jin}}]{PhysRevLett.98.200403}%
  \BibitemOpen
  \bibfield  {author} {\bibinfo {author} {\bibfnamefont {J.~P.}\ \bibnamefont
  {Gaebler}}, \bibinfo {author} {\bibfnamefont {J.~T.}\ \bibnamefont
  {Stewart}}, \bibinfo {author} {\bibfnamefont {J.~L.}\ \bibnamefont {Bohn}}, \
  and\ \bibinfo {author} {\bibfnamefont {D.~S.}\ \bibnamefont {Jin}},\ }\href
  {\doibase 10.1103/PhysRevLett.98.200403} {\bibfield  {journal} {\bibinfo
  {journal} {Phys. Rev. Lett.}\ }\textbf {\bibinfo {volume} {98}},\ \bibinfo
  {pages} {200403} (\bibinfo {year} {2007})}\BibitemShut {NoStop}%
\bibitem [{\citenamefont {Ohashi}(2005)}]{Ohashi2005BECBCS}%
  \BibitemOpen
  \bibfield  {author} {\bibinfo {author} {\bibfnamefont {Y.}~\bibnamefont
  {Ohashi}},\ }\href {\doibase 10.1103/PhysRevLett.94.050403} {\bibfield
  {journal} {\bibinfo  {journal} {Phys. Rev. Lett.}\ }\textbf {\bibinfo
  {volume} {94}},\ \bibinfo {pages} {050403} (\bibinfo {year}
  {2005})}\BibitemShut {NoStop}%
\bibitem [{\citenamefont {Peng}\ \emph {et~al.}(2014)\citenamefont {Peng},
  \citenamefont {Tan},\ and\ \citenamefont {Jiang}}]{Jiang2014pwave}%
  \BibitemOpen
  \bibfield  {author} {\bibinfo {author} {\bibfnamefont {S.-G.}\ \bibnamefont
  {Peng}}, \bibinfo {author} {\bibfnamefont {S.}~\bibnamefont {Tan}}, \ and\
  \bibinfo {author} {\bibfnamefont {K.}~\bibnamefont {Jiang}},\ }\href
  {\doibase 10.1103/PhysRevLett.112.250401} {\bibfield  {journal} {\bibinfo
  {journal} {Phys. Rev. Lett.}\ }\textbf {\bibinfo {volume} {112}},\ \bibinfo
  {pages} {250401} (\bibinfo {year} {2014})}\BibitemShut {NoStop}%
\bibitem [{\citenamefont {Luciuk}\ \emph {et~al.}(2016)\citenamefont {Luciuk},
  \citenamefont {Trotzky}, \citenamefont {Smale}, \citenamefont {Yu},
  \citenamefont {Zhang},\ and\ \citenamefont
  {Thywissen}}]{Thywissen2016pwaveContact}%
  \BibitemOpen
  \bibfield  {author} {\bibinfo {author} {\bibfnamefont {C.}~\bibnamefont
  {Luciuk}}, \bibinfo {author} {\bibfnamefont {S.}~\bibnamefont {Trotzky}},
  \bibinfo {author} {\bibfnamefont {S.}~\bibnamefont {Smale}}, \bibinfo
  {author} {\bibfnamefont {Z.}~\bibnamefont {Yu}}, \bibinfo {author}
  {\bibfnamefont {S.}~\bibnamefont {Zhang}}, \ and\ \bibinfo {author}
  {\bibfnamefont {J.~H.}\ \bibnamefont {Thywissen}},\ }\href
  {http://dx.doi.org/10.1038/nphys3670} {\bibfield  {journal} {\bibinfo
  {journal} {Nat Phys}\ }\textbf {\bibinfo {volume} {12}},\ \bibinfo {pages}
  {599} (\bibinfo {year} {2016})},\ \bibinfo {note} {article}\BibitemShut
  {NoStop}%
\bibitem [{\citenamefont {Gao}(2000)}]{Gao2000universal}%
  \BibitemOpen
  \bibfield  {author} {\bibinfo {author} {\bibfnamefont {B.}~\bibnamefont
  {Gao}},\ }\href {\doibase 10.1103/PhysRevA.62.050702} {\bibfield  {journal}
  {\bibinfo  {journal} {Phys. Rev. A}\ }\textbf {\bibinfo {volume} {62}},\
  \bibinfo {pages} {050702} (\bibinfo {year} {2000})}\BibitemShut {NoStop}%
\bibitem [{\citenamefont {Gao}(1998)}]{gao1998r6solutions}%
  \BibitemOpen
  \bibfield  {author} {\bibinfo {author} {\bibfnamefont {B.}~\bibnamefont
  {Gao}},\ }\href {\doibase 10.1103/PhysRevA.58.1728} {\bibfield  {journal}
  {\bibinfo  {journal} {Phys. Rev. A}\ }\textbf {\bibinfo {volume} {58}},\
  \bibinfo {pages} {1728} (\bibinfo {year} {1998})}\BibitemShut {NoStop}%
\bibitem [{\citenamefont {Fano}(1970)}]{Fano1970}%
  \BibitemOpen
  \bibfield  {author} {\bibinfo {author} {\bibfnamefont {U.}~\bibnamefont
  {Fano}},\ }\href {\doibase 10.1103/PhysRevA.2.353} {\bibfield  {journal}
  {\bibinfo  {journal} {Phys. Rev. A}\ }\textbf {\bibinfo {volume} {2}},\
  \bibinfo {pages} {353} (\bibinfo {year} {1970})}\BibitemShut {NoStop}%
\bibitem [{\citenamefont {Starace}(1976)}]{Starace1976}%
  \BibitemOpen
  \bibfield  {author} {\bibinfo {author} {\bibfnamefont {A.~F.}\ \bibnamefont
  {Starace}},\ }\enquote {\bibinfo {title} {The quantum defect theory
  approach},}\ in\ \href {\doibase 10.1007/978-1-4684-2799-8_23} {\emph
  {\bibinfo {booktitle} {Photoionization and Other Probes of Many - Electron
  Interactions}}},\ \bibinfo {editor} {edited by\ \bibinfo {editor}
  {\bibfnamefont {F.~J.}\ \bibnamefont {Wuilleumier}}}\ (\bibinfo  {publisher}
  {Springer US},\ \bibinfo {address} {Boston, MA},\ \bibinfo {year} {1976})\
  pp.\ \bibinfo {pages} {395--406}\BibitemShut {NoStop}%
\bibitem [{\citenamefont {Seaton}(1983)}]{Seaton1983}%
  \BibitemOpen
  \bibfield  {author} {\bibinfo {author} {\bibfnamefont {M.~J.}\ \bibnamefont
  {Seaton}},\ }\href {http://stacks.iop.org/0034-4885/46/i=2/a=002} {\bibfield
  {journal} {\bibinfo  {journal} {Reports on Progress in Physics}\ }\textbf
  {\bibinfo {volume} {46}},\ \bibinfo {pages} {167} (\bibinfo {year}
  {1983})}\BibitemShut {NoStop}%
\bibitem [{\citenamefont {Gao}(2008{\natexlab{a}})}]{Gao2008General1overRn}%
  \BibitemOpen
  \bibfield  {author} {\bibinfo {author} {\bibfnamefont {B.}~\bibnamefont
  {Gao}},\ }\href {\doibase 10.1103/PhysRevA.78.012702} {\bibfield  {journal}
  {\bibinfo  {journal} {Phys. Rev. A}\ }\textbf {\bibinfo {volume} {78}},\
  \bibinfo {pages} {012702} (\bibinfo {year} {2008}{\natexlab{a}})}\BibitemShut
  {NoStop}%
\bibitem [{\citenamefont {Mies}(1984)}]{Mies1984MQDT}%
  \BibitemOpen
  \bibfield  {author} {\bibinfo {author} {\bibfnamefont {F.~H.}\ \bibnamefont
  {Mies}},\ }\href {\doibase 10.1063/1.447000} {\bibfield  {journal} {\bibinfo
  {journal} {The Journal of Chemical Physics}\ }\textbf {\bibinfo {volume}
  {80}},\ \bibinfo {pages} {2514} (\bibinfo {year} {1984})}\BibitemShut
  {NoStop}%
\bibitem [{\citenamefont {Idziaszek}\ \emph {et~al.}(2011)\citenamefont
  {Idziaszek}, \citenamefont {Simoni}, \citenamefont {Calarco},\ and\
  \citenamefont {Julienne}}]{Idzaszek2011MQDT}%
  \BibitemOpen
  \bibfield  {author} {\bibinfo {author} {\bibfnamefont {Z.}~\bibnamefont
  {Idziaszek}}, \bibinfo {author} {\bibfnamefont {A.}~\bibnamefont {Simoni}},
  \bibinfo {author} {\bibfnamefont {T.}~\bibnamefont {Calarco}}, \ and\
  \bibinfo {author} {\bibfnamefont {P.~S.}\ \bibnamefont {Julienne}},\ }\href
  {http://stacks.iop.org/1367-2630/13/i=8/a=083005} {\bibfield  {journal}
  {\bibinfo  {journal} {New Journal of Physics}\ }\textbf {\bibinfo {volume}
  {13}},\ \bibinfo {pages} {083005} (\bibinfo {year} {2011})}\BibitemShut
  {NoStop}%
\bibitem [{\citenamefont {Moerdijk}\ \emph {et~al.}(1995)\citenamefont
  {Moerdijk}, \citenamefont {Verhaar},\ and\ \citenamefont
  {Axelsson}}]{Moerdijk1995SpinExchange}%
  \BibitemOpen
  \bibfield  {author} {\bibinfo {author} {\bibfnamefont {A.~J.}\ \bibnamefont
  {Moerdijk}}, \bibinfo {author} {\bibfnamefont {B.~J.}\ \bibnamefont
  {Verhaar}}, \ and\ \bibinfo {author} {\bibfnamefont {A.}~\bibnamefont
  {Axelsson}},\ }\href {\doibase 10.1103/PhysRevA.51.4852} {\bibfield
  {journal} {\bibinfo  {journal} {Phys. Rev. A}\ }\textbf {\bibinfo {volume}
  {51}},\ \bibinfo {pages} {4852} (\bibinfo {year} {1995})}\BibitemShut
  {NoStop}%
\bibitem [{\citenamefont {Stoof}\ \emph {et~al.}(1988)\citenamefont {Stoof},
  \citenamefont {Koelman},\ and\ \citenamefont
  {Verhaar}}]{Stoof1988SpinExchange}%
  \BibitemOpen
  \bibfield  {author} {\bibinfo {author} {\bibfnamefont {H.~T.~C.}\
  \bibnamefont {Stoof}}, \bibinfo {author} {\bibfnamefont {J.~M. V.~A.}\
  \bibnamefont {Koelman}}, \ and\ \bibinfo {author} {\bibfnamefont {B.~J.}\
  \bibnamefont {Verhaar}},\ }\href {\doibase 10.1103/PhysRevB.38.4688}
  {\bibfield  {journal} {\bibinfo  {journal} {Phys. Rev. B}\ }\textbf {\bibinfo
  {volume} {38}},\ \bibinfo {pages} {4688} (\bibinfo {year}
  {1988})}\BibitemShut {NoStop}%
\bibitem [{\citenamefont {Mies}\ \emph {et~al.}(1996)\citenamefont {Mies},
  \citenamefont {Williams}, \citenamefont {Julienne},\ and\ \citenamefont
  {Krauss}}]{Mies1996Collisions}%
  \BibitemOpen
  \bibfield  {author} {\bibinfo {author} {\bibfnamefont {F.~H.}\ \bibnamefont
  {Mies}}, \bibinfo {author} {\bibfnamefont {C.~J.}\ \bibnamefont {Williams}},
  \bibinfo {author} {\bibfnamefont {P.~S.}\ \bibnamefont {Julienne}}, \ and\
  \bibinfo {author} {\bibfnamefont {M.}~\bibnamefont {Krauss}},\ }\href
  {http://physics.nist.gov/Pubs/Bec/j4mies.pdf} {\bibfield  {journal} {\bibinfo
   {journal} {J. Res. Natl. Inst. Stand. Technol.}\ }\textbf {\bibinfo {volume}
  {101}},\ \bibinfo {pages} {521} (\bibinfo {year} {1996})}\BibitemShut
  {NoStop}%
\bibitem [{\citenamefont {Kotochigova}\ \emph {et~al.}(2000)\citenamefont
  {Kotochigova}, \citenamefont {Tiesinga},\ and\ \citenamefont
  {Julienne}}]{Kotochigova2000SecondOrderSpinOrbit}%
  \BibitemOpen
  \bibfield  {author} {\bibinfo {author} {\bibfnamefont {S.}~\bibnamefont
  {Kotochigova}}, \bibinfo {author} {\bibfnamefont {E.}~\bibnamefont
  {Tiesinga}}, \ and\ \bibinfo {author} {\bibfnamefont {P.~S.}\ \bibnamefont
  {Julienne}},\ }\href {\doibase 10.1103/PhysRevA.63.012517} {\bibfield
  {journal} {\bibinfo  {journal} {Phys. Rev. A}\ }\textbf {\bibinfo {volume}
  {63}},\ \bibinfo {pages} {012517} (\bibinfo {year} {2000})}\BibitemShut
  {NoStop}%
\bibitem [{\citenamefont {Leo}\ \emph {et~al.}(2000)\citenamefont {Leo},
  \citenamefont {Williams},\ and\ \citenamefont {Julienne}}]{Julienne2000FBCs}%
  \BibitemOpen
  \bibfield  {author} {\bibinfo {author} {\bibfnamefont {P.~J.}\ \bibnamefont
  {Leo}}, \bibinfo {author} {\bibfnamefont {C.~J.}\ \bibnamefont {Williams}}, \
  and\ \bibinfo {author} {\bibfnamefont {P.~S.}\ \bibnamefont {Julienne}},\
  }\href {\doibase 10.1103/PhysRevLett.85.2721} {\bibfield  {journal} {\bibinfo
   {journal} {Phys. Rev. Lett.}\ }\textbf {\bibinfo {volume} {85}},\ \bibinfo
  {pages} {2721} (\bibinfo {year} {2000})}\BibitemShut {NoStop}%
\bibitem [{\citenamefont {Hanna}\ \emph {et~al.}(2009)\citenamefont {Hanna},
  \citenamefont {Tiesinga},\ and\ \citenamefont
  {Julienne}}]{PhysRevA.79.040701}%
  \BibitemOpen
  \bibfield  {author} {\bibinfo {author} {\bibfnamefont {T.~M.}\ \bibnamefont
  {Hanna}}, \bibinfo {author} {\bibfnamefont {E.}~\bibnamefont {Tiesinga}}, \
  and\ \bibinfo {author} {\bibfnamefont {P.~S.}\ \bibnamefont {Julienne}},\
  }\href {\doibase 10.1103/PhysRevA.79.040701} {\bibfield  {journal} {\bibinfo
  {journal} {Phys. Rev. A}\ }\textbf {\bibinfo {volume} {79}},\ \bibinfo
  {pages} {040701} (\bibinfo {year} {2009})}\BibitemShut {NoStop}%
\bibitem [{\citenamefont {Strauss}\ \emph {et~al.}(2010)\citenamefont
  {Strauss}, \citenamefont {Takekoshi}, \citenamefont {Lang}, \citenamefont
  {Winkler}, \citenamefont {Grimm}, \citenamefont {Hecker~Denschlag},\ and\
  \citenamefont {Tiemann}}]{PhysRevA.82.052514}%
  \BibitemOpen
  \bibfield  {author} {\bibinfo {author} {\bibfnamefont {C.}~\bibnamefont
  {Strauss}}, \bibinfo {author} {\bibfnamefont {T.}~\bibnamefont {Takekoshi}},
  \bibinfo {author} {\bibfnamefont {F.}~\bibnamefont {Lang}}, \bibinfo {author}
  {\bibfnamefont {K.}~\bibnamefont {Winkler}}, \bibinfo {author} {\bibfnamefont
  {R.}~\bibnamefont {Grimm}}, \bibinfo {author} {\bibfnamefont
  {J.}~\bibnamefont {Hecker~Denschlag}}, \ and\ \bibinfo {author}
  {\bibfnamefont {E.}~\bibnamefont {Tiemann}},\ }\href {\doibase
  10.1103/PhysRevA.82.052514} {\bibfield  {journal} {\bibinfo  {journal} {Phys.
  Rev. A}\ }\textbf {\bibinfo {volume} {82}},\ \bibinfo {pages} {052514}
  (\bibinfo {year} {2010})}\BibitemShut {NoStop}%
\bibitem [{\citenamefont {Gao}(2009)}]{gao2009analytic}%
  \BibitemOpen
  \bibfield  {author} {\bibinfo {author} {\bibfnamefont {B.}~\bibnamefont
  {Gao}},\ }\href {\doibase 10.1103/PhysRevA.80.012702} {\bibfield  {journal}
  {\bibinfo  {journal} {Phys. Rev. A}\ }\textbf {\bibinfo {volume} {80}},\
  \bibinfo {pages} {012702} (\bibinfo {year} {2009})}\BibitemShut {NoStop}%
\bibitem [{\citenamefont {K\"ohler}\ \emph {et~al.}(2006)\citenamefont
  {K\"ohler}, \citenamefont {G\'oral},\ and\ \citenamefont
  {Julienne}}]{Julienne2006RMP}%
  \BibitemOpen
  \bibfield  {author} {\bibinfo {author} {\bibfnamefont {T.}~\bibnamefont
  {K\"ohler}}, \bibinfo {author} {\bibfnamefont {K.}~\bibnamefont {G\'oral}}, \
  and\ \bibinfo {author} {\bibfnamefont {P.~S.}\ \bibnamefont {Julienne}},\
  }\href {\doibase 10.1103/RevModPhys.78.1311} {\bibfield  {journal} {\bibinfo
  {journal} {Rev. Mod. Phys.}\ }\textbf {\bibinfo {volume} {78}},\ \bibinfo
  {pages} {1311} (\bibinfo {year} {2006})}\BibitemShut {NoStop}%
\bibitem [{\citenamefont {Ticknor}\ \emph {et~al.}(2004)\citenamefont
  {Ticknor}, \citenamefont {Regal}, \citenamefont {Jin},\ and\ \citenamefont
  {Bohn}}]{ticknor2004multiplet}%
  \BibitemOpen
  \bibfield  {author} {\bibinfo {author} {\bibfnamefont {C.}~\bibnamefont
  {Ticknor}}, \bibinfo {author} {\bibfnamefont {C.~A.}\ \bibnamefont {Regal}},
  \bibinfo {author} {\bibfnamefont {D.~S.}\ \bibnamefont {Jin}}, \ and\
  \bibinfo {author} {\bibfnamefont {J.~L.}\ \bibnamefont {Bohn}},\ }\href
  {\doibase 10.1103/PhysRevA.69.042712} {\bibfield  {journal} {\bibinfo
  {journal} {Phys. Rev. A}\ }\textbf {\bibinfo {volume} {69}},\ \bibinfo
  {pages} {042712} (\bibinfo {year} {2004})}\BibitemShut {NoStop}%
\bibitem [{\citenamefont {Peik}\ \emph {et~al.}(1997)\citenamefont {Peik},
  \citenamefont {Ben~Dahan}, \citenamefont {Bouchoule}, \citenamefont
  {Castin},\ and\ \citenamefont {Salomon}}]{peik1997bloch}%
  \BibitemOpen
  \bibfield  {author} {\bibinfo {author} {\bibfnamefont {E.}~\bibnamefont
  {Peik}}, \bibinfo {author} {\bibfnamefont {M.}~\bibnamefont {Ben~Dahan}},
  \bibinfo {author} {\bibfnamefont {I.}~\bibnamefont {Bouchoule}}, \bibinfo
  {author} {\bibfnamefont {Y.}~\bibnamefont {Castin}}, \ and\ \bibinfo {author}
  {\bibfnamefont {C.}~\bibnamefont {Salomon}},\ }\href {\doibase
  10.1103/PhysRevA.55.2989} {\bibfield  {journal} {\bibinfo  {journal} {Phys.
  Rev. A}\ }\textbf {\bibinfo {volume} {55}},\ \bibinfo {pages} {2989}
  (\bibinfo {year} {1997})}\BibitemShut {NoStop}%
\bibitem [{\citenamefont {Marte}\ \emph {et~al.}(2002)\citenamefont {Marte},
  \citenamefont {Volz}, \citenamefont {Schuster}, \citenamefont {D\"urr},
  \citenamefont {Rempe}, \citenamefont {van Kempen},\ and\ \citenamefont
  {Verhaar}}]{PhysRevLett.89.283202}%
  \BibitemOpen
  \bibfield  {author} {\bibinfo {author} {\bibfnamefont {A.}~\bibnamefont
  {Marte}}, \bibinfo {author} {\bibfnamefont {T.}~\bibnamefont {Volz}},
  \bibinfo {author} {\bibfnamefont {J.}~\bibnamefont {Schuster}}, \bibinfo
  {author} {\bibfnamefont {S.}~\bibnamefont {D\"urr}}, \bibinfo {author}
  {\bibfnamefont {G.}~\bibnamefont {Rempe}}, \bibinfo {author} {\bibfnamefont
  {E.~G.~M.}\ \bibnamefont {van Kempen}}, \ and\ \bibinfo {author}
  {\bibfnamefont {B.~J.}\ \bibnamefont {Verhaar}},\ }\href {\doibase
  10.1103/PhysRevLett.89.283202} {\bibfield  {journal} {\bibinfo  {journal}
  {Phys. Rev. Lett.}\ }\textbf {\bibinfo {volume} {89}},\ \bibinfo {pages}
  {283202} (\bibinfo {year} {2002})}\BibitemShut {NoStop}%
\bibitem [{\citenamefont {Schuster}\ \emph {et~al.}(2012)\citenamefont
  {Schuster}, \citenamefont {Scelle}, \citenamefont {Trautmann}, \citenamefont
  {Knoop}, \citenamefont {Oberthaler}, \citenamefont {Haverhals}, \citenamefont
  {Goosen}, \citenamefont {Kokkelmans},\ and\ \citenamefont
  {Tiemann}}]{schuster2012feshbach}%
  \BibitemOpen
  \bibfield  {author} {\bibinfo {author} {\bibfnamefont {T.}~\bibnamefont
  {Schuster}}, \bibinfo {author} {\bibfnamefont {R.}~\bibnamefont {Scelle}},
  \bibinfo {author} {\bibfnamefont {A.}~\bibnamefont {Trautmann}}, \bibinfo
  {author} {\bibfnamefont {S.}~\bibnamefont {Knoop}}, \bibinfo {author}
  {\bibfnamefont {M.~K.}\ \bibnamefont {Oberthaler}}, \bibinfo {author}
  {\bibfnamefont {M.~M.}\ \bibnamefont {Haverhals}}, \bibinfo {author}
  {\bibfnamefont {M.~R.}\ \bibnamefont {Goosen}}, \bibinfo {author}
  {\bibfnamefont {S.~J. J. M.~F.}\ \bibnamefont {Kokkelmans}}, \ and\ \bibinfo
  {author} {\bibfnamefont {E.}~\bibnamefont {Tiemann}},\ }\href {\doibase
  10.1103/PhysRevA.85.042721} {\bibfield  {journal} {\bibinfo  {journal} {Phys.
  Rev. A}\ }\textbf {\bibinfo {volume} {85}},\ \bibinfo {pages} {042721}
  (\bibinfo {year} {2012})}\BibitemShut {NoStop}%
\bibitem [{\citenamefont {T~Pack}\ \emph {et~al.}(1998)\citenamefont {T~Pack},
  \citenamefont {Walker},\ and\ \citenamefont {Kendrick}}]{Pack98}%
  \BibitemOpen
  \bibfield  {author} {\bibinfo {author} {\bibfnamefont {R.}~\bibnamefont
  {T~Pack}}, \bibinfo {author} {\bibfnamefont {R.~B.}\ \bibnamefont {Walker}},
  \ and\ \bibinfo {author} {\bibfnamefont {B.~K.}\ \bibnamefont {Kendrick}},\
  }\href
  {http://scitation.aip.org/content/aip/journal/jcp/109/16/10.1063/1.477348}
  {\bibfield  {journal} {\bibinfo  {journal} {The Journal of Chemical Physics}\
  }\textbf {\bibinfo {volume} {109}},\ \bibinfo {pages} {6701} (\bibinfo {year}
  {1998})}\BibitemShut {NoStop}%
\bibitem [{\citenamefont {Forrey}(2015)}]{Forrey15}%
  \BibitemOpen
  \bibfield  {author} {\bibinfo {author} {\bibfnamefont {R.~C.}\ \bibnamefont
  {Forrey}},\ }\href
  {http://scitation.aip.org/content/aip/journal/jcp/143/2/10.1063/1.4926325}
  {\bibfield  {journal} {\bibinfo  {journal} {The Journal of Chemical Physics}\
  }\textbf {\bibinfo {volume} {143}},\ \bibinfo {eid} {024101} (\bibinfo {year}
  {2015})}\BibitemShut {NoStop}%
\bibitem [{\citenamefont {Gao}()}]{Gaoupb}%
  \BibitemOpen
  \bibfield  {author} {\bibinfo {author} {\bibfnamefont {B.}~\bibnamefont
  {Gao}},\ }\href@noop {} {}\bibinfo {note} {Unpublished}\BibitemShut {NoStop}%
\bibitem [{\citenamefont {Fu}\ \emph {et~al.}(2016)\citenamefont {Fu},
  \citenamefont {Li}, \citenamefont {Tey}, \citenamefont {You},\ and\
  \citenamefont {Gao}}]{Fu16}%
  \BibitemOpen
  \bibfield  {author} {\bibinfo {author} {\bibfnamefont {H.}~\bibnamefont
  {Fu}}, \bibinfo {author} {\bibfnamefont {M.}~\bibnamefont {Li}}, \bibinfo
  {author} {\bibfnamefont {M.~K.}\ \bibnamefont {Tey}}, \bibinfo {author}
  {\bibfnamefont {L.}~\bibnamefont {You}}, \ and\ \bibinfo {author}
  {\bibfnamefont {B.}~\bibnamefont {Gao}},\ }\href
  {http://stacks.iop.org/1367-2630/18/i=10/a=103016} {\bibfield  {journal}
  {\bibinfo  {journal} {New Journal of Physics}\ }\textbf {\bibinfo {volume}
  {18}},\ \bibinfo {pages} {103016} (\bibinfo {year} {2016})}\BibitemShut
  {NoStop}%
\end{thebibliography}
%

\end{document}